# Three-dimensional imaging of individual carbon atoms

Na Yeon Kim[1*], Hanfeng Zhong[1,2*], Jianhua Zhang[3*], Colum M. O'Leary[1*], Yuxuan Liao[1], Ji Zou[4], Haozhi Sha[1], Minh Pham[1], Weiyi Li[1], Yakun Yuan[1], Ji-Hoon Park[5], Dennis Kim[6], Huaidong Jiang[3], Jing Kong[5], Miaofang Chi[7], Jianwei Miao[1†]

[1]Department of Physics and Astronomy and California NanoSystems Institute, University of California, Los Angeles, CA, USA.
[2]Department of Electrical and Computer Engineering, University of California, Los Angeles, CA, USA.
[3]Center for Transformative Science, ShanghaiTech University, Shanghai, China.
[4]Department of Physics, University of Basel, Basel, Switzerland.
[5]Department of Electrical Engineering and Computer Science, Massachusetts Institute of Technology, Cambridge, MA, USA.
[6]Department of Chemistry and Biochemistry, University of California, Los Angeles, CA, USA.
[7]Center for Nanophase Materials Sciences, Oak Ridge National Laboratory, Oak Ridge, TN, USA.
[*]These authors contributed equally to this work. [†]Correspondence author. Email: j.miao@ucla.edu

**Carbon is fundamental to science and technology due to its diverse bonding configurations, structural versatility, and essential role in defining the mechanical, chemical, electronic, and quantum properties of materials. However, direct three-dimensional (3D) imaging of individual carbon atoms remains a long-standing challenge. Here, we use twisted bilayer graphene (TBG) as a model system and demonstrate ptychographic atomic electron tomography (pAET) for determining the 3D atomic coordinates of individual carbon atoms with a precision of 0.11 Å. The resulting 3D atomic model uncovers chiral lattice distortions driven by van der Waals interactions that exhibit meron-like and skyrmion-like structural textures. These findings provide direct insight into the interplay between 3D chiral lattice deformation and electronic properties in moiré-engineered carbon systems. Beyond TBG, pAET offers a versatile approach for 3D atomic-scale imaging of carbon-based and other light-element materials that are central to advances in physics, chemistry, materials science, and nanotechnology.**

Carbon's unique bonding versatility gives rise to a wide range of structural forms, including diamond, graphite, amorphous carbon, fullerenes, carbon nanotubes, and graphene (*1-3*). In these materials, atomic-scale distortions such as lattice corrugation, interlayer misalignment, strain, disorder, and point defects can significantly influence mechanical, thermal, electronic, and optical properties by altering local bonding environments and structural order (*4-8*). However, conventional imaging techniques, including scanning probe microscopy (*8-10*), transmission electron microscopy (*11-15*), and scanning transmission electron microscopy (*16-18*), are generally constrained to surface sensitivity or 2D projections, limiting access to 3D atomic structure. This challenge is particularly critical in carbon-based systems, where even subtle 3D displacements can strongly affect material properties. As a result, accurately determining the 3D coordinates of individual carbon atoms is essential for elucidating structure-property relationships.

Ptychography, a computational microscopy technique that unifies crystallography with microscopy by combining coherent diffraction and iterative phase retrieval, replaces lenses with computation and enables quantitative, aberration-free imaging of both light and heavy elements across multiple modalities (*19*). In particular, multislice electron ptychography exploits the curvature of diffracted waves to retrieve depth information without sample tilting or through-focus series (*20*, *21*) and has been applied to 3D imaging of light-element materials (*22-25*). However, the achievable depth resolution remains constrained by the convergence and diffraction angles and by the inherent challenge of reconstructing high-resolution 3D structures from a single viewing direction (*20*, *24*). Combining small sample tilts with multislice ptychography can improve depth resolution to about 9 Å (*26*), but this remains insufficient to localize individual light atoms in three dimensions. To determine 3D atomic structures without assuming crystallinity, atomic electron tomography (AET) has proven to be the most effective approach (*27*). Annular dark-field AET has achieved picometer-scale 3D atomic structure determination in heavy-element materials (*28–30*). Recently, ptychographic AET (pAET) has been proposed and experimentally tested, but determining the 3D atomic positions of low-Z materials such as carbon ($Z \leqslant 8$) remains an outstanding challenge (*31-34*). Here, we demonstrate that pAET overcomes these limitations, enabling high-precision 3D atomic structure determination in carbon-based materials. Our pAET workflow (fig. S1), which integrates data acquisition, ptychographic phase retrieval, projection alignment, tomographic reconstruction, atom tracing, position refinement, and structural analysis, achieves 3D imaging of individual carbon atoms with 11 pm precision.

**pAET experiment of TBG**

We chose TBG as a model system due to its significance as a quantum material (*35*) and the necessity to resolve 3D atomic positions for deeper insight into interlayer interactions within moiré patterns (*36*). The TBG sample was prepared by stacking graphene monolayers at a small twist angle (*37*). A focused electron probe was used for single-atom fabrication, introducing individual silicon dopants at defect sites in TBG (fig. S2) (*37*). These dopants induced local strain and served as reference points for tomographic alignment. The experiment was performed using a 70 keV electron beam, with the TBG sample tilted to 13 different angles using a double-tilt stage (*37*). To minimize vertical vibrations of the freestanding TBG, regions near the edge of the support hole were selected. At each tilt angle, diffraction patterns were recorded by scanning the sample with a 70 keV focused electron probe in a 2D raster pattern (Fig. 1A and table S1). These diffraction patterns encode electron-sample interactions, providing input for computational phase retrieval algorithms that reconstruct the complex functions of both the sample and the probe (*19*). By capturing nearly all transmitted and diffracted electrons, ptychography enhances the signal-to-noise ratio and resolution, overcoming the limitations of conventional electron microscopy (*38*).

Reconstruction was performed using the multimode extended ptychographic iterative engine (ePIE) (*39*), incorporating four orthogonal probe modes to account for partial coherence and measurement imperfections (fig. S3 and S4) (*40*). The reconstructed phase images of TBG at 13 tilt angles, together with the corresponding annular dark-field images, are shown in fig. S5, S6, and movie S1, where individual silicon dopants are visible and their positions remain consistent across all projections. High-precision tomographic reconstruction required accurate tilt angle calibration, which was achieved using nanoparticles as fiducial markers in low-magnification annular dark-field images (fig. S7), acquired after each ptychography measurement at every tilt angle. The calibrated angles were refined and combined with silicon dopants to align the 13 ptychographic projections (*37*). To ensure consistency, the phase images were scaled to maintain a uniform mean phase value across all projections (*41*).

After image pre-processing, the 13 ptychographic projections and their calibrated tilt angles were used to reconstruct the 3D atomic structure of TBG (*37*). Tomographic reconstruction was performed using a real-space iterative algorithm, and convergence was quantitatively monitored by the R-factor (*42*). To mitigate the effects of the limited tilt range, we incorporated a 3D axial probe into the tomographic reconstruction algorithm (*37*). During each iteration, the 3D object was divided into multiple slices along the projection direction, with each slice assigned a defocus-dependent probe. Compared with the simplified parallel-beam approximation, this approach provides a more accurate physical model and improves depth resolution by compensating for defocus variations introduced at different tilt angles (Fig. 1B). From the 3D reconstruction, we quantified the resolution by selecting 862 carbon atoms, performing background subtraction, and averaging their subvolumes to obtain a statistically representative 3D atomic profile. By fitting a 3D Gaussian function to this profile, we measured resolutions of 1.01 Å, 0.73 Å, and 2.94 Å along the x, y, and z axes, respectively (fig. S8), with the lower z-axis resolution arising from the limited experimental tilt range.

By projecting the 3D reconstruction along the z axis, we obtained a 1D intensity profile that revealed two distinct peaks corresponding to the top and bottom graphene layers (Fig. 1B, blue curve). This profile was used to define a 1D support constraint along the z axis, modeled as a pair of generalized Gaussian distributions and applied together with the axial probe during tomographic reconstruction. Varying the kernel size of the generalized Gaussian functions yielded consistent 3D reconstructions (Fig. 1, C and D). Quantitative analysis shows that an optimally tuned 1D z-support improved the 3D resolution to 1.09, 0.76, and 1.49 Å along the x, y, and z axes, respectively (fig. S9), with a significant enhancement along the z axis that increased the 3D precision in identifying individual atoms (fig. S10). Figure 1, E and F, shows the top and bottom layers of TBG, clearly separated following the incorporation of the axial probe and 1D z-support constraint into the tomographic reconstruction (Fig. 1G). This separation is further supported by the Fourier transforms of the two layers, which reveal a twist angle of 1.97°, indicating minimal crosstalk between them (fig. S11).

**Resolving individual 3D atomic positions with picometer-level precision**

To identify individual atoms with improved precision despite the limited number of tilt angles and the presence of a large missing wedge, we developed an iterative atom-tracing algorithm to extract 3D atomic coordinates from the tomographic reconstruction (*37*). Unlike previous non-iterative approaches (*28-30*), our algorithm identifies atomic positions iteratively, with tracing errors converging after just a few iterations (fig. S12). The resulting 3D atomic coordinates were further refined using a gradient descent optimization. Each atom was represented by a 3D Gaussian function that accounts for the electron probe size and experimental uncertainties. For each experimental projection, calculated projections were generated over a range of tilt angles centered on the calibrated values (fig. S7B). The 3D atomic coordinates, tilt angles, and projection alignment were iteratively refined by minimizing the difference between the experimental and calculated projections (37). The refined parameters were then used for the final tomographic reconstruction with the axial probe and 1D z-support, followed by iterative atom tracing and refinement to obtain the final 3D atomic model. To test robustness, we introduced random positional and angular shifts into the experimental projections. The pAET workflow automatically corrected these misalignments, demonstrating consistent convergence of both projection and angular refinements (figs. S13 and S14).

To quantify the 3D precision of pAET, we performed multislice simulations to generate 13 sets of ptychographic diffraction patterns based on the experimental 3D atomic model and refined tilt angles, under the same experimental conditions used in data acquisition (table S1) (*37*). Thermal vibrations were modeled by incorporating 64 frozen phonons. Poisson noise was added to the simulated diffraction patterns to match the experimental dose conditions, and the 13 ptychographic projections were reconstructed using the multimode ePIE algorithm (*39, 40*). To account for broadening effects arising from partial coherence of the electron probe and other experimental factors, each projection was convolved with a Gaussian kernel (*43*). The kernel size was optimized to minimize discrepancies between the simulated and experimental phase projections. These 13 simulated projections were then processed using the same pAET workflow (fig. S1) as applied to the experimental data, producing a validation atomic model.

Figure 2, A to D, compares the 3D atomic positions between the experimental and validation models, yielding root mean square deviations (RMSDs) of 6, 5, and 7 pm along the x, y, and z axes, respectively, and an overall RMSD of 11 pm. The result was further supported by multislice simulations using a randomly perturbed version of the DFT-relaxed atomic model, which, when processed through the same pAET workflow (fig. S1), produced similar RMSD values (figs. S15 and S16). To quantify the detectability limits, we introduced random atomic displacements of 10, 15, 20, and 30 pm into simulated TBG datasets generated under experimental conditions. The method detected 10 pm displacements with a true-positive rate of 76.9% and a false-positive rate of 17.2%, with the accuracy further improving at larger displacements of 15, 20, and 30 pm (*37*). A likelihood-ratio test comparing the experimental projections with those calculated with and without a 10 pm out-of-plane (z-axis) shift yielded $\Delta\chi^2 = 32.5$ ($p = 1.2 \times 10^{-8}$), corresponding to a Gaussian-equivalent significance of 5.7σ (*37*), confirming that such a displacement is unambiguously detectable in the experimental data.

Figure 2, E and F, shows the top and bottom layers of the experimental 3D atomic model, which contains 6,649 carbon atoms and 3 silicon dopants (fig. S17). Yellow and blue regions represent upward and downward atomic corrugations, respectively, while black lines indicate $sp^2$ C-C bonds. The bond length was determined to be 1.44 ± 0.02 Å (Fig. 2G), closely matching the DFT-relaxed model of 1.425 Å (fig. S18). High-precision experimental measurements enabled a quantitative analysis of interlayer spacing as a function of local stacking configuration. The overall average interlayer distance in the experimental model is 3.43 ± 0.03 Å (mean ± SD), consistent with the DFT-relaxed model value of 3.41 ± 0.04 Å (fig. S19). However, a more detailed comparison reveals notable discrepancies at the local level: in the experimental model, the interlayer spacing is 3.45 ± 0.02 Å in AA regions and 3.41 ± 0.03 Å in AB/BA regions (Fig. 2H), whereas the DFT model yields corresponding values of 3.49 ± 0.01 Å and 3.37 ± 0.006 Å. These differences underscore the greater complexity of atomic corrugations in real TBG compared to theoretical predictions. This is further corroborated by distinct structural

variations observed in the AA stacking regions: in one domain (AA1, Fig. 2, E and F), both layers curve upward, while in another (AA2), the layers exhibit a pronounced bulging deformation. Such intricate atomic-scale corrugations can induce charge redistribution, disrupt local rotational symmetry, and alter the electronic band structure of TBG (*44*).

**Quantifying 3D chiral lattice distortions in TBG**

Using the experimental 3D atomic model, we quantitatively analyzed 3D displacement vectors in TBG, which represents a significant advancement over previous experimental efforts limited to 2D characterization (*45, 46*). To enhance the statistical precision of the displacement vector field, the atomic positions were convolved with a 3D Gaussian kernel whose widths corresponded to the in-plane lattice constant and the average interlayer spacing along the z-axis. This smoothing suppresses high-frequency noise and local fluctuations, enabling more robust extraction of long-wavelength structural distortions. To investigate the displacement field driven by van der Waals interactions, we compared the experimental 3D atomic model with an unrelaxed TBG model composed of two flat graphene layers (*37*). This comparison enabled extraction of 3D displacement vectors for both the top and bottom layers (Fig. 3, A and B). The displacement vectors show strong in-plane vorticity around the AA stacking regions, accompanied by out-of-plane components along the z axis, which exhibit chiral lattice distortions closely resembling magnetic merons and skyrmions (*47*).

To quantify these features, we calculated the solid angle subtended by each chiral configuration, defined as the skyrmion number (*37*). The experimental TBG model reveals two meron-like, three anti-meron-like, and one anti-skyrmion-like structural textures. Figure 3, C and D, illustrates the 3D displacement vector fields of a meron-like and an anti-meron-like structural textures, both featuring elliptical geometries with skyrmion numbers of -0.42 and 0.46, respectively. Additional meron-like and anti-meron-like structural textures are shown in fig. S20. The displacement vectors are oriented along the z axis near the core, with their polar angles gradually increasing outward, tilting predominantly into the xy plane at the boundary. The experimentally observed vector fields deviate from those in the DFT-relaxed TBG model, which display perfectly symmetric meron-like and anti-meron-like structural textures with skyrmion numbers of -0.5 and +0.5, respectively (fig. S21, A and B). Figure 3E shows the 3D displacement vector field of an anti-skyrmion-like structural texture, characterized by an elliptical geometry and a skyrmion number of 0.97. This configuration features an anti-meron-like core (Fig. 3D), with some boundary vectors tilting their polar angles close to 180°. Compared with its counterpart in the DFT-relaxed model (fig. S21C), the experimental anti-skyrmion-like structural textures exhibit deviations in geometry, skyrmion number, and the configuration of its 3D displacement field.

To assess the robustness of the chiral lattice distortions observed in TBG, we first examined the effect of Gaussian kernel size by repeating the analysis with kernel widths reduced by half along the x, y, and z axes. Although the smaller kernel increases local fluctuations, the main features of the 3D displacement vector field and the associated chiral lattice distortions remain consistent between the two cases (Fig. 3 and fig. S22). We further tested robustness by performing leave-one-tilt-out pAET reconstructions for all 13 projections, resulting in 13 independent 3D atomic models. Figure S23 shows the 3D displacement fields and chiral lattice distortions in the top and bottom graphene layers from a representative leave-one-tilt-out reconstruction, which are in good agreement with those in Fig. 3, A and B.

To evaluate the effect of the support constraint on the 3D displacement vector fields, we reconstructed the 13 experimental projections using a pair of transient square functions as a 1D z-support. Each function constrained a single graphene layer (width 1.52 Å) and was applied only during the first 50 iterations, then lifted for the subsequent 150 iterations (*37*). The resulting 3D atomic model was subsequently refined against the experimental projections without any support constraint. The RMSD between this model and that obtained using the generalized Gaussian support is 8 pm, and the corresponding 3D displacement vector fields show strong agreement (fig. S24 vs. Fig. 3, A and B). These results demonstrate that appropriate use of the z-support, followed by support-free refinement, does not introduce artificial structural features in TBG.

Building on this structural analysis, we investigated how chiral lattice distortions affect the electronic properties of TBG by incorporating experimentally determined 3D atomic coordinates into DFT calculations. Although DFT has been widely used to predict TBG's electronic properties (*48*), it typically relies on idealized lattice models with manually introduced defects, which are then relaxed to their ground-state configurations. By directly inputting experimentally measured atomic coordinates, our approach captures the intrinsic structural complexity of real TBG and enables a more accurate calculation of its electronic band structure. Figure 4A shows a moiré supercell containing 3,267 carbon atoms and a silicon dopant, including two anti-meron-like structural textures. Compared to the DFT-relaxed model, the experimental 3D atomic coordinates exhibit larger out-of-plane variations along the z axis (Fig. 4, B and C). Figure 4D compares the electronic band structure derived from the experimental 3D atomic coordinates with that of the DFT-relaxed model. In the relaxed model, the valence and conduction bands meet at the K point near the Fermi level (green), forming a Dirac cone. This feature indicates that relaxed TBG behaves as a semimetal, analogous to monolayer graphene (*49*).

In contrast, the experimental moiré supercell exhibits a Dirac cone with a 22 meV bandgap (blue). At room temperature, the thermal energy (~26 meV) is comparable to the bandgap, allowing thermal excitation of charge carriers. As the temperature decreases, the carrier population drops exponentially, reducing conductivity and rendering TBG increasingly insulating at lower temperatures. Moreover, the larger out-of-plane atomic displacements in the experimental model raise the valence band and lower the conduction band energies at the Γ point. This results in a bandwidth reduction of up to 13.5%, indicating stronger interlayer electronic coupling in real TBG.

## Conclusions

We demonstrate that pAET achieves picometer-precision determination of the 3D coordinates of individual carbon atoms in TBG, revealing interlayer spacing variations governed by local stacking. Quantitative analysis of the 3D displacement vector field provides, to our knowledge, the first experimental evidence of chiral lattice distortions driven by van der Waals interactions, including meron-like, anti-meron-like, and anti-skyrmion-like structural textures. Incorporating the experimentally determined 3D atomic coordinates into DFT calculations reveals an asymmetric electronic band structure with a 22 meV bandgap, in contrast to the zero-bandgap Dirac cone predicted by the DFT-relaxed model. Beyond TBG, we demonstrate the general applicability of the pAET workflow (fig. S1) through multislice simulations that resolve a nitrogen-vacancy center embedded deep within a 15 nm diamond crystal (fig. S25) and determine the 3D atomic structure of a 5 nm amorphous carbon nanoparticle (*37*; fig. S26). With further optimization, pAET should enable sub-angstrom 3D imaging across a wide range of low-Z materials, including van der Waals heterostructures (*36*), quantum emitters (*6*), superconductors (*8*, *25*), charge-density-wave systems (*50*), ferroelectrics (*51*), catalysts (*52*), and amorphous carbon (3), thereby providing atomic-scale insight into structure–function relationships in complex and beam-sensitive systems.

**ACKNOWLEDGMENTS**

We acknowledge Juan Carlos Idrobo and Andrew R. Lupini for their assistance with the experiments, and Yao Yang for help with data analysis. This research was primarily supported by the U.S. Department of Energy, Office of Science, Basic Energy Sciences, Division of Materials Sciences and Engineering, under award no. DE-SC0010378. N.Y.K., C.O.L., J.-H.P., and J.K. acknowledge support from the U.S. Army Research Office MURI project under grant W911NF-18-1-04320431. H.Z. acknowledges support from the U.S. Air Force Office of Scientific Research MURI project under Award No. FA9550-23-1-0281. J.Z. thanks support from the Georg H. Endress Foundation. Electron ptychography and STEM experiments were conducted at the Center for Nanophase Materials Sciences, Oak Ridge National Laboratory. Multislice simulations were performed at NERSC, a U.S. Department of Energy Office of Science User Facility, under awards BES-ERCAP0028035 and ERCAP0034513.

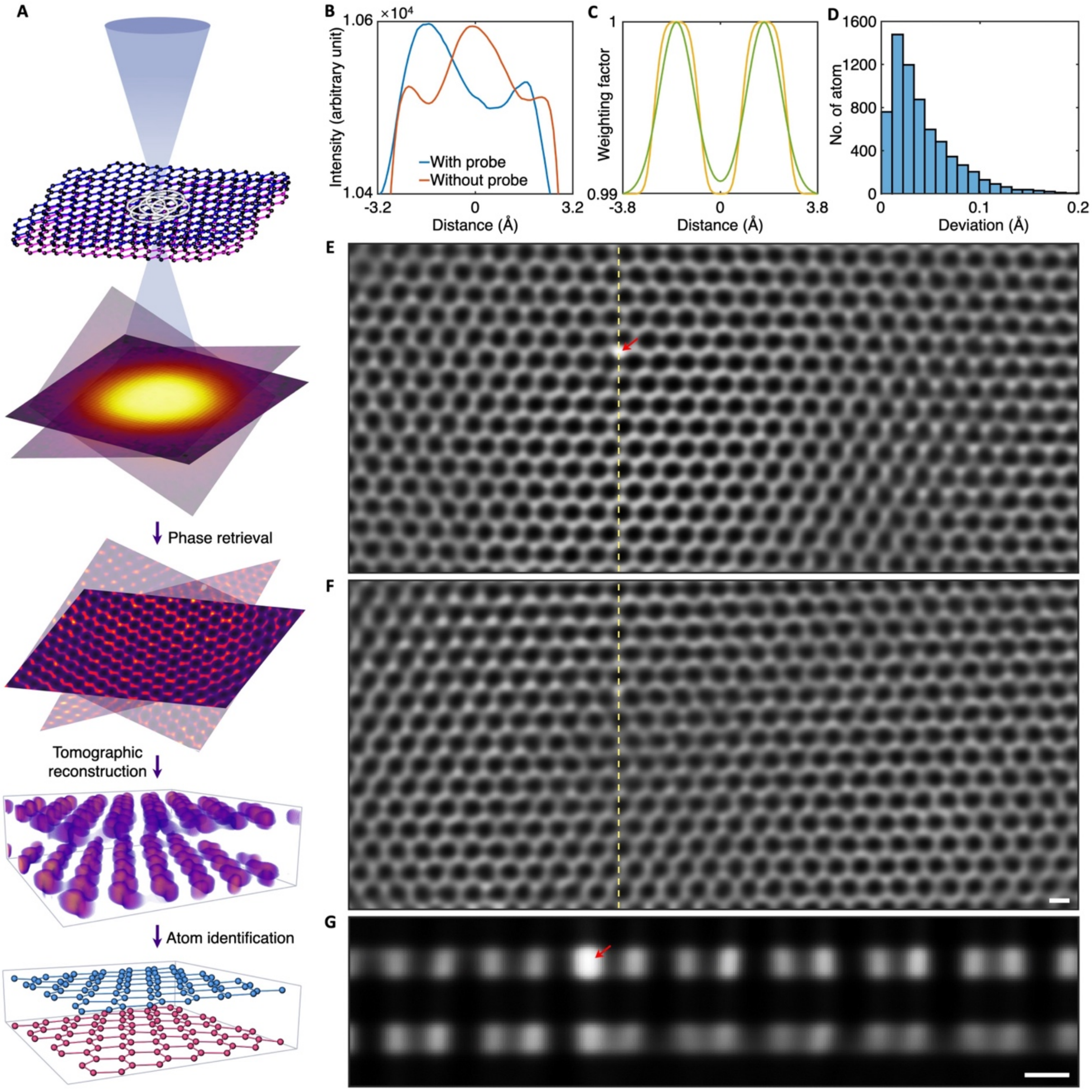

**Fig. 1. Experimental demonstration of pAET for resolving the 3D coordinates of individual carbon atoms in TBG.** (**A**) Schematic of the pAET workflow. A focused electron probe scans the sample, producing diffraction patterns that are reconstructed into phase projections. These projections serve as input for tomographic reconstruction, followed by atom tracing and position refinement. (**B**) Incorporation of a 3D axial probe into the tomographic reconstruction improves depth resolution and layer separation. (**C**) Application of a 1D support constraint along the z axis, modeled as a pair of generalized Gaussian distributions with different kernel sizes (yellow and green), iteratively applied to enhance structural precision. (**D**) Comparison of tomographic reconstructions using different kernel sizes, yielding an RMSD of 5 pm. (**E**, **F**) Top (E) and bottom (F) graphene layers resolved after applying both the 3D axial probe kernel and the 1D support constraint. Red arrows indicate silicon dopants. (**G**) Projection along the z axis showing both layers and the region outlined in (E) and (F), confirming effective layer separation. Scale bars, 2 Å.

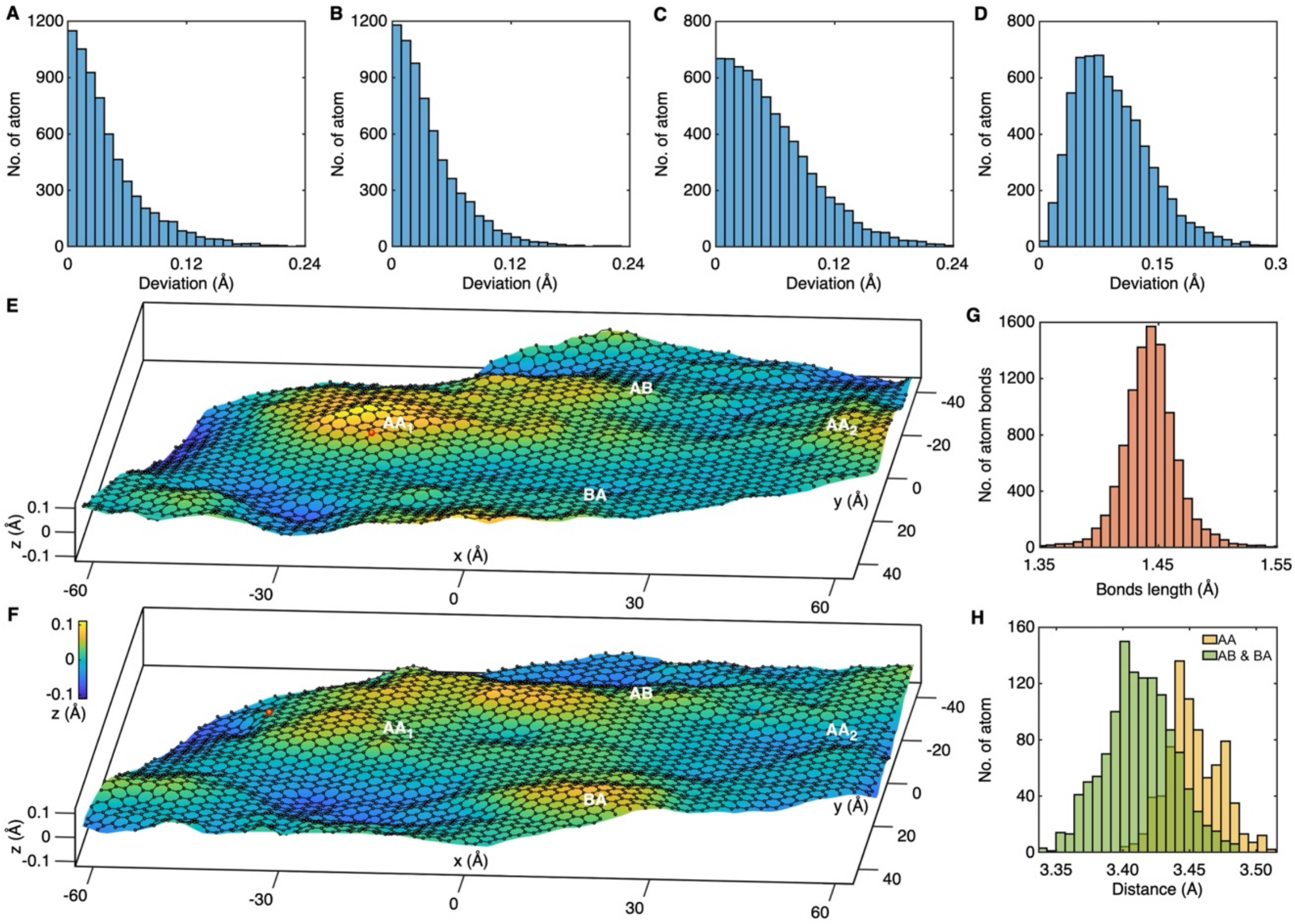

**Fig. 2. 3D atomic corrugations in experimental TBG arising from stacking-dependent interlayer interactions.** (**A-D**) Histograms of atomic position deviations between the experimental and validation models along the x, y, and z axes, and overall, yielding RMSDs of 6, 5, 7, and 11 pm, respectively. (**E**, **F**) Top (E) and bottom (F) graphene layers showing out-of-plane corrugations. Yellow and blue indicate upward and downward curvature, respectively. To improve displacement field precision at the expense of resolution, atomic positions were convolved with a 3D Gaussian kernel with standard deviations of 1.26 Å in x and y (half the in-plane lattice constant) and 1.71 Å in z (half the average interlayer spacing). (**G**) Histogram of $sp^2$ C–C bond lengths in the experimental model, showing a mean of 1.44 Å with a standard deviation of 0.02 Å. (**H**) Histogram of interlayer distances in AA and AB/BA stacking regions, with mean values of 3.45 Å (SD = 0.02 Å) and 3.41 Å (SD = 0.03 Å), respectively.

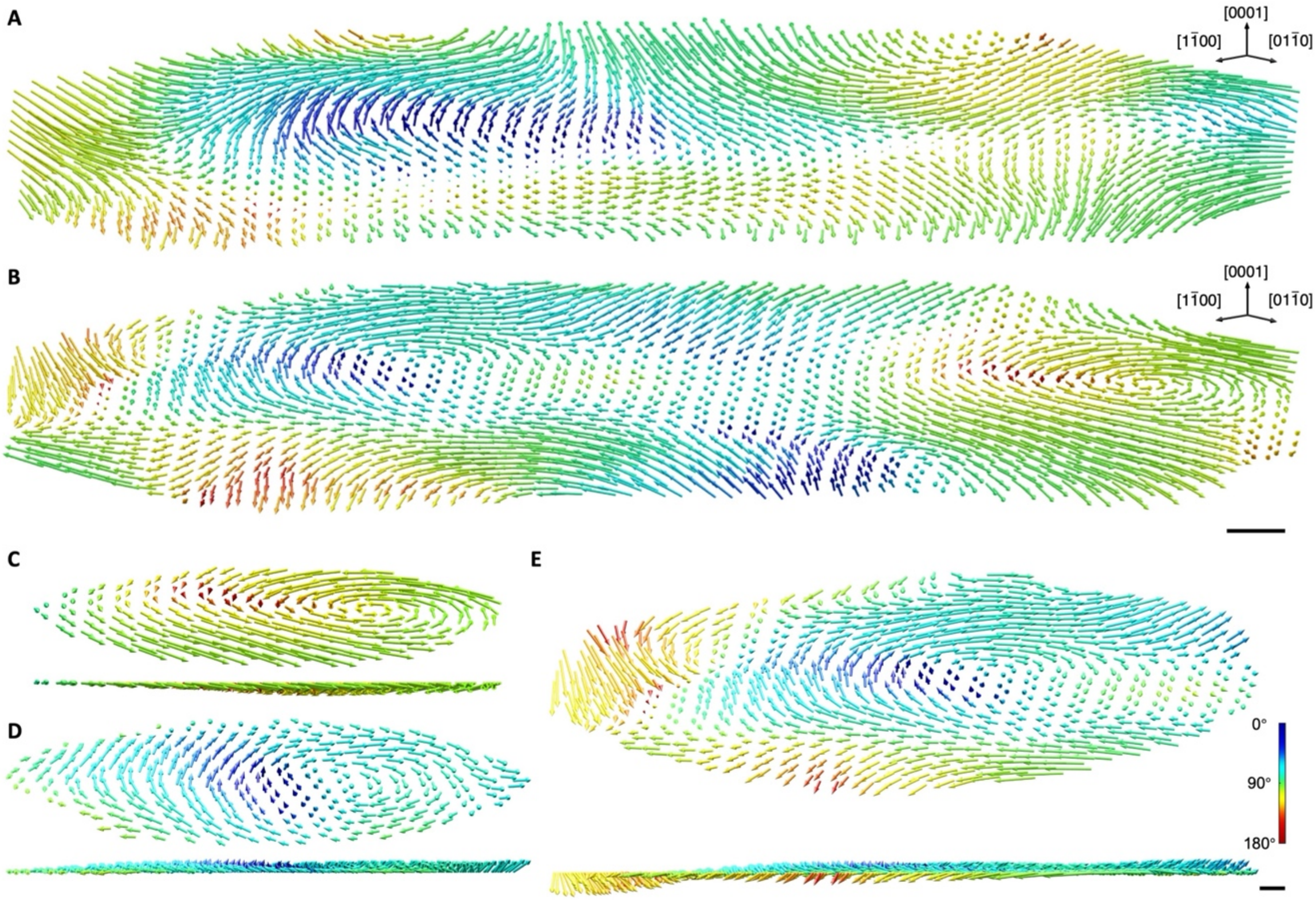

**Fig. 3. Quantification of 3D displacement vectors and chiral lattice distortions in experimental TBG.** (**A**, **B**) 3D displacement vector fields in the top and bottom graphene layers. Vector color represents the polar angle relative to the xy plane. (**C**, **D**) Localized 3D vector fields of a meron-like (C) and an anti-meron-like (D) structural textures. Near the core, vectors align with the z axis and gradually tilt into the xy plane toward the boundaries, characteristic of chiral lattice distortions. (**E**) 3D vector field of an anti-skyrmion-like structural texture incorporating the anti-meron-like core from (D). Several boundary vectors approach polar angles near 180°, indicating more complex out-of-plane deformation. Vector magnitudes are scaled by a factor of 25 for visualization. Scale bar, 5 Å in (B) and 2 Å in (E).

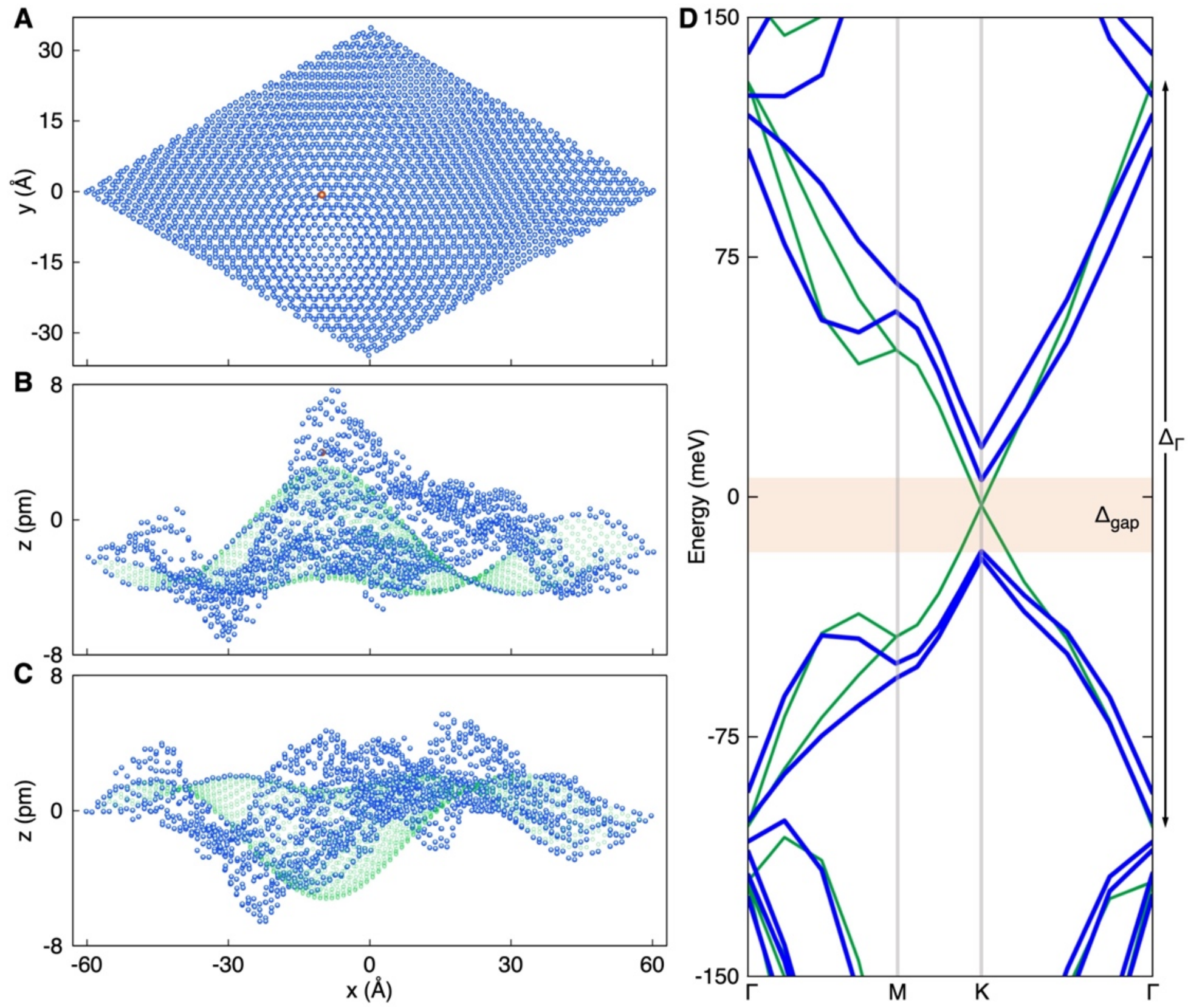

**Fig. 4. Comparison of electronic band structures in TBG from experimental and DFT-relaxed atomic models.** (**A**) Experimental 3D atomic structure of a moiré supercell containing 3,267 carbon atoms and one silicon dopant (red dot), including two anti-meron-like features. (**B**, **C**) Overlay of atomic positions from the experimental (blue) and DFT-relaxed (green) models in the top (B) and bottom (C) graphene layers, showing greater out-of-plane (z-axis) deviations in the experimental structure, indicative of enhanced corrugation. (**D**) Electronic band structures computed from the two models. The experimental model (blue) exhibits an asymmetric band structure with a 22 meV gap at the K point and a bandwidth reduction of up to 13.5% at the Γ point. The DFT-relaxed model (green) retains a symmetric band structure with a gapless Dirac cone.

## Materials and Methods

TBG synthesis

Monolayer graphene was synthesized via chemical vapor deposition (*53*) (CVD) on a 25 µm thick Cu foil (Alfa Aesar, 99.8%, #13382). The CVD-grown graphene was spin-coated with Poly(methyl methacrylate) (PMMA), followed by selective etching of the Cu foil using ferric chloride solution (CE-100, Transene Inc.). The graphene/PMMA layer was rinsed multiple times in DI water to remove Cu etchant residues and transferred onto a TEM sample grid (Quantifoil™ 200 mesh Au R 1.2/1.3). After air drying, the grid was heated to 90 °C to improve adhesion between the graphene/PMMA layer and the grid. The PMMA layer was then removed by soaking in a 60 °C acetone bath, and the graphene/grid was rinsed in isopropyl alcohol. To eliminate volatile organics, the grid was annealed under $Ar/H_2$ gas (300 sccm/100 sccm) at 350 °C for 3 hours. A second graphene layer was transferred onto the grid using the same procedure, resulting in twisted bilayer graphene. To minimize electron-beam-induced amorphous carbon contamination during data collection, UVO treatment (Zone-TEM, Hitachi Ltd.) was applied to both sides of the sample grid for 5 minutes. Finally, the sample and holder stage were annealed in vacuum ($10^{-3}$ Pa, 160 °C) for 18 hours to remove volatile compounds before data acquisition.

Electron-beam manipulation of individual dopant atoms

To incorporate individual silicon dopants into TBG lattices, a 100-keV converged electron beam was raster-scanned over regions of TBG enriched with dopant sources (fig. S2). The incident electrons, exceeding the carbon knock-on threshold energy (*54*), created atomic vacancies in the TBG lattice. Subsequently, the electron beam was precisely applied in line-profile scanning to transfer source atoms from impurity-rich areas to these defect sites. The dopants formed stable $sp^3$ bonds within the carbon lattice, serving as fiducial markers for tomographic alignment.

Ptychography data acquisition

Electron ptychography data were collected on a Nion UltraSTEM 100 scanning transmission electron microscope (STEM) equipped with an optically coupled Hamamatsu Orca CMOS detector (2048 × 2048 pixels) and a double-tilt side-entry holder. The holder provided two orthogonal tilt axes, each with a nominal range of ±22°. The detector was windowed to 48 × 48 pixels with a dwell time of 1.5 ms per probe position. An accelerating voltage of 70 kV was used to minimize knock-on damage to carbon, and a convergence semi-angle of 32 mrad was selected to achieve atomic resolution with high contrast. To optimize the imaging conditions, we performed preliminary tests and found that a focused probe yielded slightly better reconstruction quality for our specific setup. A probe current of 60 pA and a scan step size of 0.65 Å were employed, yielding 230 × 230 diffraction patterns over the TBG region of interest. The electron dose per tilt was $1.3 \times 10^6$ $e^-$ Å$^{-2}$, giving a total dose of $1.69 \times 10^7$ $e^-$ Å$^{-2}$, which is below the threshold for defect formation in TBG (*55*). Data acquisition was repeated at 13 tilt angles using a double-tilt holder, with the specific angles listed in fig. S7B. To mitigate beam effects during acquisition, we introduced ~1 h intervals between tilts to allow thermal and mechanical relaxation and to reduce holder hysteresis. Beam blanking was applied between scans, and dwell times were minimized to limit radiolysis and contamination.

Ptychographic reconstruction and image pre-processing

Each TBG projection was reconstructed using a customized multimode ePIE algorithm (*39*). Diffraction patterns were zero-padded to 192 × 192 pixels, yielding a real-space pixel size of 0.095 Å. To account for partial coherence, four orthogonal probe modes were used in the reconstruction (*40*, *56*) (figs. S3 and S4). Robustness was evaluated by repeating the reconstructions with two, four, and six modes, with and without a detector response model; results with four or more modes were highly consistent and insensitive to the detector model. Both single- and multi-slice ptychographic reconstructions produced very similar results. Figure S5 shows the 13 ptychographic phase images across the tilt series. Phase images were aligned using the center-of-mass method (*57*), with Si substitutional dopants serving as fiducial markers.

Angle calibration

The nominal angles were measured using the Nion double tilt holder during data acquisition but required calibration to account for stage hysteresis. To calibrate the angles, we utilized low-magnification annular dark-field images featuring multiple metal nanoparticles. The x and y coordinates of each nanoparticle were determined through 2D Gaussian fitting. The tilt angles of the 13 projections were calibrated using the least-squares method, minimizing the error function $E$,

$$E = \sum_i \sum_j \left\{ \left[ P_x(r_i, \alpha_j, \beta_j) - x_i^j \right]^2 + \left[ P_y(r_i, \alpha_j, \beta_j) - y_i^j \right]^2 \right\} \qquad (1)$$

where $P_x$ and $P_y$ are the functions projecting the 3D coordinates of nanoparticles onto the x and y coordinates, respectively. $r_i$ denotes the 3D coordinates of the $i^{\text{th}}$ nanoparticle, while $\alpha_j$ and $\beta_j$ are the tilt angles for the $j^{\text{th}}$ projection. The measured x and y coordinates for the $i^{\text{th}}$ particle in the $j^{\text{th}}$ projection are given by $x_i^j$ and $y_i^j$. Utilizing Equation (1), we determined the 3D coordinates of all nanoparticles within the field of view for each tilt projection, which allowed us to calculate the calibrated tilt angles and tilt axes (fig. S7). The least-squares fitting yielded consistent results across different initial inputs.

Tomographic reconstruction incorporating a 3D axial probe kernel

The pre-processed dataset was used as input to an enhanced version of the REal Space Iterative REconstruction (RESIRE) algorithm (*42*). This algorithm iteratively minimizes the least-squares difference between the ptychographic projection and the corresponding backprojection from the 3D reconstruction. To account for the probe's depth dependence, we incorporated an axial probe kernel $K(z)$ into the loss function $\varepsilon_\theta$ for each projection angle $\theta$,

$$\varepsilon_\theta(O) = \frac{1}{2} \sum_{x,y} \left| \sum_z K(z) * O_\theta\{x,y,z\} - b_\theta\{x,y\} \right|^2 \qquad (2)$$

where $*$ denotes the convolution operator, $O_\theta\{x,y,z\}$ represents the rotation of $O\{x,y,z\}$ relative to the tilt angle $\theta$, and $b_\theta$ is the corresponding measured projection. The axial probe kernel was defined from the experimental probe parameters (electron voltage and convergence semi-angle) and positioned relative to the sample according to the measured defocus. Because the projection operators in Eq. (2) are linear, we solved the minimization problem using gradient descent. Quantitative simulations demonstrated that this kernel produced reconstructions of higher fidelity than those obtained with either the parallel-beam assumption or a more complex probe function extracted from ptychographic reconstructions. In the tomographic reconstruction of our experimental data, the number of slices varied with the tilt angle: higher tilt angles required more slices, while lower tilt angles required fewer. For the 13 projections, we used between 5 and 15 slices, with each slice having a thickness of 80 pixels (7.6 Å). Reconstructions performed with different slice thicknesses yielded consistent atomic coordinates, confirming that the results are robust to the choice of slice sampling.

For the 13 experimental projections, we performed three iterative reconstruction cycles using the RESIRE algorithm. The initial cycle consisted of 80 iterations, producing a preliminary 3D reconstruction, which was then used to derive a pair of generalized Gaussian distributions with varying kernel sizes as a 1D support in the z-direction (see the following Methods section). Next, RESIRE was run for 200 iterations, incorporating the 3D electron probe and the 1D support, to generate an enhanced 3D reconstruction. Spatial alignment and angular refinement were then conducted by aligning the experimental projections with the calculated projections (fig. S7). Finally, a third RESIRE cycle consisting of 200 iterations, incorporating both the axial probe and the 1D support, was performed to obtain the refined 3D reconstruction. For each tilt, we calculated the R-factor between the experimental projection and the corresponding projection from the 3D reconstruction. The 13 values are 11.9%, 9.3%, 12.5%, 13.5%, 11.0%, 12.7%, 10.1%, 9.9%, 12.2%, 11.1%, 14.4%, 10.1%, and 9.7%, with a mean of 11.4%. These values fall well within the range reported in previous AET studies (*28-30*), indicating that the dataset is self-consistent and that no systematic beam-induced damage occurred during the tilt series.

Estimation of the 3D resolution

To quantify the 3D resolution, we selected 862 carbon atoms from the tomographic reconstruction obtained without applying a z-axis support constraint. To suppress low-frequency background variations, a background subtraction was first performed. For each selected atom, a 3D subvolume was extracted, aligned to the atomic center, and averaged to generate a representative single-atom profile. This averaging was used solely for resolution estimation. Orthogonal slices through the averaged volume in the xy, xz, and yz planes are shown in Fig. S8, A–C. A 3D Gaussian function was fitted to the averaged atomic profile, and the full width at half maximum (FWHM) along the x, y, and z axes was used to determine the spatial resolution. The measured FWHMs were 1.01 Å (x), 0.73 Å (y), and 2.94 Å (z), respectively (Fig. S8, D–F). The in-plane resolution is sufficient to resolve individual carbon atoms, while the axial resolution enables clear separation of the 3.43 Å interlayer spacing in TBG. We also quantified the 3D resolution of the experimental reconstruction after applying the 1D z-support. Figure S9, A–C, shows orthogonal slices in the xy, xz, and yz planes from a 3D volume obtained by statistically averaging the same 862 carbon atoms. Fitting a 3D Gaussian to the averaged atom yielded FWHMs of 1.09 Å (x), 0.76 Å (y), and 1.49 Å (z), demonstrating that the 1D z-support constraint substantially improves the z-axis resolution.

Utilizing a pair of generalized Gaussian distributions with varying kernel sizes as 1D supports

Applying a 1D support constraint using a pair of generalized Gaussian distributions (*58*) effectively separates the individual graphene monolayers along the z-direction in the tomographic reconstruction. Initially, RESIRE was integrated with the 3D electron probe to identify two peaks along the z-direction by summing the intensities across all projections. From the two separated peaks (Fig. 1B, blue curve), we applied a pair of generalized Gaussian distributions (*58*) with varying kernel sizes to generate 1D supports along the z-axis. The kernel size (β) determines the extent of the support region along the z-direction; a larger β results in a broader generalized Gaussian distribution. To assess the impact of different β values on 3D precision, we performed three independent tomographic reconstructions incorporating the electron probe and 1D supports with β = 2, 3 and 4, where β = 2 corresponds to a Gaussian function. From the three independent reconstructions, we traced the 3D atomic coordinates of all atoms and calculated the RMSD. The RMSD was 5 pm between β = 2 and β = 3, and 5.4 pm between β = 2 and 4 (Fig. 1, C and D). The small RMSDs demonstrate the robustness of incorporating a 1D support constraint. For the final 3D reconstruction shown in Fig. 1, E to G, we used a kernel width parameter of β = 4 to achieve clear separation between the top and bottom graphene layers while minimizing the constraint's influence on structural detail. This separation is further supported by the Fourier transforms of the individual layers (fig. S11), which exhibit non-overlapping diffraction features, confirming effective spatial separation.

<u>Iterative atom tracing from the final tomographic reconstruction</u>

To address the limited number of projections and the presence of a large missing wedge, we developed an iterative atom-tracing method to identify the 3D atomic coordinates from the final tomographic reconstruction. Figure S12A shows a diagram of the method, which began tracing atoms from the final reconstruction using the polynomial method (*28, 30, 59*), while enforcing a minimum nearest-neighbour distance of 1.35 Å. After examining all the local minima, a set of 3D coordinates for the initial atoms was obtained, from which an average initial atom with dimensions of 2.19 Å × 2.19 Å × 2.19 Å was calculated. The final reconstruction was subtracted by the average initial atom at each traced atom position, and the polynomial method was applied to trace the residual atoms, from which an average residual atom was generated. This average residual atom, along with the traced residual atom positions, was used for the next iteration. At each iteration, an $R_1$ factor was calculated, defined as the difference between the calculated and experimental projections (*28*). The calculated projections were derived from a 3D structure in which all traced atom positions were filled with an average atom. Compared to previous non-iterative atom-tracing approaches (*28-30*), the current method refines atom positions by iterating between initial and residual atoms. Figure S12B shows the decrease in the $R_1$ factor as the iteration number increases, with the process terminating after 10 iterations. Since not all atoms were initially identified, the intermediate reconstruction was corrected by subtracting an average atomic density at the traced atom positions from the 10th iteration, followed by additional tracing using the polynomial method. The iterative automated tracing algorithm successfully identified approximately 93% of the atoms. The remaining atoms were located by combining manual inspection with automated tracing. Specifically, a 3D Gaussian function was fitted to each local maximum to determine the centroids of the remaining ~7% of atoms. In total, 6,649 carbon atoms and 3 silicon atoms were identified.

<u>Refining the 3D atomic coordinates</u>

The traced 3D atomic coordinates were refined by minimizing the error between the calculated and experimental images using gradient descent. Specifically, each atom was modelled as a 3D Gaussian function to generate a 3D atomic model (*28*),

$$O^{\{x,y,z\}} = \sum_i H_i \exp\left( -\frac{|x - x_i|^2 + |y - y_i|^2 + |z - z_i|^2}{B_i^{'}} \right) \quad (3)$$

where $x_i$, $y_i$, and $z_i$ represent the coordinates of the $i^{th}$ atom, $H_i$ is the scaling factor, and $B_i'$ accounts for the electron probe size, thermal motions, and experimental error of the $i^{th}$ atom. The fitted 3D atomic model was then linearly projected to generate images using the experimental tilt angles. To improve precision, positional and angular refinement was performed using the 3D atomic model. For each experimental projection, a series of calculated projections was generated from the model over a range of tilt angles centered around the experimentally calibrated values. The tilt angles were then refined by maximizing the cross-correlation between the calculated and experimental projections. This process was iteratively repeated until no further improvement was detected, yielding a refined set of atomic positions and optimized tilt angles for the experimental projections (fig. S7B). The final $R_1$ factor of the atom refinement process is 12%. To eliminate edge effects, atoms located at the boundary of the atomic model were excluded from the position refinement.

3D precision estimation

To estimate the 3D precision of our pAET method, we simulated ptychographic diffraction patterns using abTEM (*60*). A total of 13 datasets of diffraction patterns were generated, matching the experimental parameters (table S1). Thermal vibrations of atomic positions were accounted for by incorporating 64 frozen phonons into the multislice method. After adding Poisson noise to the diffraction patterns, ptychographic reconstruction was performed for each tilt angle using the multimode ePIE algorithm (*39, 40*). Each ptychographic phase image was then convolved with a Gaussian kernel to account for broadening effects from the electron probe and sample drift. Using the same RESIRE reconstruction along with the atomic tracing and refinement procedures applied to the experimental data, we generated a new 3D atomic model. A quantitative comparison between this model and the experimental one yielded RMSDs of 6, 5, and 7 pm along the x, y, and z axes, respectively, with an overall RMSD of 11 pm (Fig. 2, A to D).

To provide a more rigorous estimate of our method's precision, we implemented a randomized validation framework by applying uniformly distributed random displacements with a maximum amplitude of ±10 pm along the x, y, and z axes to the DFT-relaxed atomic model of TBG (fig. S15). Multislice simulations were then performed on the perturbed model to generate electron ptychography datasets, using the same tilt geometry and imaging parameters as in the experimental acquisition. To account for thermal vibrations, we incorporated 32 frozen phonons. These datasets were processed using the same pAET workflow as applied to the experimental data, including ptychographic reconstruction, tomographic reconstruction with a z-axis support constraint, iterative atom tracing, and position refinement, resulting in a reconstructed atomic model. Figure S15 compares the atomic coordinates of the ground-truth and reconstructed models. Quantitative analysis yields RMSD of 4 pm, 5 pm, and 8 pm along the x, y, and z axes, respectively, corresponding to an overall RMSD of 10 pm (fig. S16). This 3D precision is consistent with the 11 pm overall RMSD obtained using the experimental 3D atomic model as input and incorporating 64 frozen phonons in the multislice simulation (Fig. 1D).

Atomic displacement detectability analysis

To quantify the detectability limits of pAET, we introduced controlled atomic displacements into the TBG structure. Random displacements of 10 pm were applied to approximately half of the atoms in a selected subset of the experimental 3D model to generate a modified structure. Diffraction patterns were then simulated under experimental conditions using multislice calculations with added Poisson noise. The noisy diffraction patterns were processed through the complete pAET workflow, including ptychographic reconstruction, convolution of each projection with a Gaussian kernel to account for partial coherence, tomographic reconstruction, atom tracing, and refinement. Comparison between the reconstructed and original models produced a histogram of atomic displacements. K-means clustering of the two peaks yielded a true-positive rate of 76.9% and a false-positive rate of 17.2% for 10 pm displacements. Repeating the analysis for displacements of 15, 20, and 30 pm produced true-positive/false-positive rates of 88.7%/11.0%, 92.5%/3.7%, and 94.5%/2.4%, respectively, demonstrating that pAET can reliably detect atomic displacements as small as 10 pm, with accuracy improving at larger amplitudes.

Likelihood-ratio ($\Delta\chi^2$) test on experimental data

We performed a likelihood-ratio test to evaluate the detectability of a 10 pm out-of-plane displacement in the experimental data. The analysis was carried out on a cropped region containing 1,853 atoms. Model $H_0$ represents the reference structure with no imposed displacement, while model $H_1$ is identical except for a 10 pm shift of the top graphene layer along the z axis. For each model, we generated diffraction patterns using multislice simulations under the same experimental conditions, added Poisson noise, and reconstructed the 13 phase projections using the same ptychography pipeline employed for the experimental data. To account for partial coherence and other broadening effects, each projection was convolved with a Gaussian kernel. Edge regions were masked to exclude boundary artifacts. We then computed the Poisson ($\chi^2$-equivalent) log-likelihoods of the 13 experimental projections under both models and summed over all pixels and tilt angles. The total $\Delta\chi^2 = \chi^2(H_0) - \chi^2(H_1) = 32.5$, decisively favoring $H_1$. With one effective degree of freedom, this corresponds to a p-value of $p = 1.2 \times 10^{-8}$, equivalent to a Gaussian-equivalent significance of approximately 5.7 σ. These results demonstrate that a 10 pm out-of-plane displacement of the top graphene layer is unambiguously detectable in the experimental data, confirming the sensitivity and statistical reliability of the pAET pipeline.

Determination of 3D atomic displacement vectors

To calculate atomic displacement vectors in TBG, we first created an unrelaxed TBG model consisting of two flat graphene layers with an interlayer distance of 3.43 Å, consistent with the average interlayer distance observed in the experiment. The atomic displacement vectors were then determined by calculating the differences in atomic positions between the experimental model and the rigid model. To enhance the statistical precision of the displacement vector field, the atomic positions were convolved with 3D Gaussian kernels. This smoothing suppresses high-frequency noise and local fluctuations, enabling more robust extraction of long-wavelength structural distortions. Two kernel sizes were applied: one with standard deviations equal to half the lattice constant in the xy plane ($\sigma_x = \sigma_y = 1.26$ Å) and half the average interlayer distance along z ($\sigma_z = 1.71$ Å), and another with standard deviations equal to the full lattice constant in the xy plane and the full interlayer distance along z.

Figure 3 shows the 3D displacement vectors and chiral lattice distortions in the experimental TBG structure after convolution with the larger Gaussian kernel. For comparison, fig. S22 presents the same analysis using the smaller kernel. Although convolution with the larger kernel reduces local fluctuations and produces a smoother displacement field, the key features of the 3D displacement vectors and chiral lattice distortions remain consistent between the two analyses. To test the robustness of the observed displacement fields, we performed multislice simulations of a flat bilayer graphene model with a 1.97° twist angle and an interlayer spacing of 3.43 Å. The simulated diffraction patterns were processed using the same pAET pipeline (fig. S1) applied to the experimental data, yielding a reconstructed atomic model. After convolving this model with both Gaussian kernels, no spurious structural textures were observed, confirming that the observed chiral distortions in the experimental data are intrinsic and not artifacts of the analysis procedure.

Calculation of the skyrmion number

To compute the skyrmion number, we normalized the displacement vectors, triangulated the TBG lattices, and calculated the solid angle (Ω) of each triangular surface subtended by three normalized vectors ($n_1$, $n_2$, and $n_3$). The solid angle is given by (*61, 62*)

$$tan\frac{\Omega}{2} = \frac{n_1 \cdot (n_2 \times n_3)}{1 + n_1 \cdot n_2 + n_1 \cdot n_3 + n_2 \cdot n_3} \quad (4)$$

The skyrmion number was calculated by summing the skyrmion densities of all triangles within the region,

$$N_{ls} = \frac{1}{4\pi} \sum_{all\ triangles} \Omega \quad (5)$$

Validation of 1D z-support preserving 3D displacement vector fields

To evaluate whether the support constraint influences the measured displacement field, we performed an independent tomographic reconstruction of the 13 experimental projections using a pair of transient square functions as a 1D z-support. Each square function constrained one graphene layer along z (width 1.52 Å, larger than any plausible in-plane variation). During the first 50 iterations, voxel values outside the support were set to zero, while those inside remained unchanged. For the subsequent 150 iterations, the support constraint was lifted, allowing the iterative reconstruction to evolve freely without any imposed prior. The resulting 3D atomic model, obtained from the tomographic reconstruction, was then refined against the experimental projections without any support constraint, ensuring that the refinement process was unbiased. The RMSD between this model and that obtained with a generalized Gaussian z-support was 16 pm before refinement and 8 pm after refinement. The 3D displacement vector fields reconstructed with the transient square support closely match those obtained with the generalized Gaussian support (fig. S24 vs. Fig. 3, A and B). Together, these results demonstrate that, when used appropriately and followed by support-free refinement, the 1D z-support does not alter the 3D displacement fields or the chiral lattice textures in TBG.

DFT calculations of relaxed and experimental 3D atomic models of a moiré supercell

A moiré supercell of TBG was constructed using the following approach. Let $\boldsymbol{a}_i$ for $i = 1, 2$ represent the lattice vector that define the primitive cell of the top layer in the TBG. The lattice vector of the bottom layer is given by $\boldsymbol{b}_i = R(\theta)\boldsymbol{a}_i$, where $R(\theta)$ is the rotation matrix and $\theta$ is the twist angle. The lattice vectors of the moiré supercell are defined as, $\boldsymbol{L}_1 = m\boldsymbol{a}_1 + n\boldsymbol{a}_2$ and $\boldsymbol{L}_2 = -R\left(\frac{\pi}{3}\right)\boldsymbol{L}_1$, where $m$ and $n$ are integers. The twist angle $\theta$ and the lattice constant $L$ of the moiré supercell can be expressed in terms of the integers $m$ and $n$,

$$\cos\theta = \frac{m^2+n^2+4mn}{2(m^2+n^2+mn)}, \quad L = |\boldsymbol{L}_1| = |\boldsymbol{L}_2| = \frac{a|m-n|}{2\sin\left(\frac{\theta}{2}\right)} \tag{6}$$

where $a = |\boldsymbol{a}_1| = |\boldsymbol{a}_2|$ is the lattice constant of the primitive cell (*63*). When $L$ equals the moiré pattern period $L_m = \frac{a}{2\sin\left(\frac{\theta}{2}\right)}$, the condition $n - m = 1$ holds. A twist angle $\theta$ = 2.005° corresponds to the pair $(n, m)$ = (17, 16). Under this condition, the superlattice constant $L$ = 7.01 nm, which is consistent with the measured moiré pattern period of approximately 7 nm. After determining the values of $n$ and $m$, the TBG moiré supercell can be constructed, comprising 3,268 atoms. The supercell dimension along the z axis was set to 12 Å, which is sufficient for convergence according to prior benchmarks (*64*).

The constructed TBG moiré supercell was relaxed using the Vienna Ab initio Simulation Package (*65*). For the DFT calculations, we used the rev-vdW-DF2 exchange–correlation functional to account for van der Waals interactions (*66*), as previous studies have reported that rev-vdW-DF2 yields bond lengths and interlayer distances in close agreement with experimental measurements (*64*). A plane-wave basis set with a cutoff energy of 420 eV was used to ensure convergence. The global convergence criterion for the self-consistent electronic loop was set to $1.0 \times 10^{-4}$ eV, and the maximum residual force on each atom was relaxed to below $1.0 \times 10^{-2}$ eV Å$^{-1}$. To relate this force tolerance to an effective geometric (corrugation) resolution, we used stiffness values extracted from the curvature of the rev-vdW-DF2 energy surface ($k_{x,y} \approx 11$ eV Å$^{-2}$ and $k_z \approx 0.093$ eV Å$^{-2}$) (*67*) to convert residual forces into displacements, obtaining < 1 pm in x and y and ~2.6 pm in z on average. These residual displacements smaller than the experimental per-atom precisions of 5, 6, and 7 pm along x, y, and z, respectively, confirming that the $1.0 \times 10^{-2}$ eV Å$^{-1}$ force threshold is sufficient for reliable comparison between the experimental and DFT-relaxed structures. Brillouin-zone sampling employed a Γ-centered 2 × 2 × 1 k-point mesh, which is adequate because the moiré supercell enlarges the real-space period by ~30× and correspondingly reduces the Brillouin-zone volume. K-point convergence was verified by the agreement of charge densities obtained from the 2 × 2 × 1 and 1 × 1 × 1 meshes. For band-structure calculations, five k-points were used along each high-symmetry path (Γ–M, M–K, K–Γ).

To derive the electronic band structure from the experimental data, we constructed a moiré supercell using the experimentally measured 3D atomic coordinates of TBG, comprising 3,267 carbon atoms and one silicon atom. Owing to the large size of the supercell, our DFT calculations indicate that the presence of a single silicon atom has a negligible effect on the overall band structure. To enhance precision at the expense of resolution, the atomic coordinates were convolved with a 3D Gaussian kernel with standard deviations of 1.26 Å along x and y (half the lattice constant) and 1.71 Å along z (half the average interlayer spacing). To minimize boundary artifacts, the edges of the experimental supercell were replaced with those from the relaxed model. The band structure of the experimental TBG was then calculated along the high-symmetry path Γ–K–M–Γ within the first Brillouin zone, using the same potential and exchange–correlation functional as in the relaxed model.

By comparing the relaxed and experimental 3D atomic models of the full moiré supercell, our DFT calculations (Fig. 4) revealed differences in the electronic band structure arising from sample-specific corrugation, stacking registry, and local strain in TBG (Fig. 4). Because our primary objective was to compare band structures derived from experimental and relaxed models, the relative differences are more meaningful than the absolute values. As with all semilocal and van der Waals–corrected DFT approaches, the absolute band gaps are lower bounds in the absence of quasiparticle corrections. For reference, semilocal DFT calculations typically underestimate band gaps of conventional semiconductors by about 30% relative to experiment, illustrating the expected level of systematic error in absolute gap values (*68*).

Multislice pAET simulations of a diamond nanocrystal containing a nitrogen-vacancy center

A 3D atomic model of a diamond nanocrystal was constructed with 12 × 12 × 43 unit cells. To create a nitrogen-vacancy (NV) center, one carbon atom was removed from the central unit cell, and a neighbouring carbon site was replaced with a substitutional nitrogen atom. The atomic coordinates of the 3 × 3 × 3 unit cells surrounding the NV center were relaxed using Vienna Ab initio Simulation Package (*65*), employing the local density approximation based on the projector-augmented-wave method (*69, 70*). The 3D atomic coordinates were used as input to abTEM (*60*) to generate ptychographic datasets with the following parameters: accelerating voltage of 70 kV, convergence semi-angle of 32 mrad, collection angle of 70 mrad, 84×84 pixels per diffraction pattern, and a scan step size of 0.4 Å (table S1). To account for thermal diffuse scattering observed in experiments, diffraction patterns were simulated using 16 frozen phonons for the projected potential. Poisson noise was added to the diffraction patterns to simulate a finite electron dose of $1.0 \times 10^4$ $e^- Å^{-2}$ per projection. Due to the 15 nm thickness of the diamond nanocrystal, a large angular tilt range was required to achieve an accurate tomographic reconstruction. A total of 57 tilt angles, ranging from -70° to 70° in 2.5° increments, were used to simulate ptychographic data along a single tilt axis.

Single- and multislice ptychographic reconstructions were performed using a maximum likelihood algorithm implemented in fold-slice, a MATLAB package adapted from Ptychoshelves (*71-73*). Unlike ePIE (*39*), fold-slice includes a regularization term on the object function that suppresses unphysical phase fluctuations along the depth, thereby enhancing the stability of multislice ptychographic reconstructions (*24, 25, 72, 73*). This algorithm improves noise tolerance and enhances reconstruction accuracy, particularly in datasets with limited redundancy. Diffraction patterns were padded from 84×84 to 128×128 pixels prior to reconstruction to achieve a real-space pixel size of 0.22 Å. Each reconstruction was performed with 200 iterations using a single probe mode. The optimization step sizes for both the object and probes were set to 0.1. In the multislice ptychographic reconstruction, the object function was divided into 20 slices, each 1 nm thick, with a regularization factor of 0.5.

Using two sets of 57 projections obtained through single- and multislice ptychography, we conducted tomographic reconstructions with and without incorporating the electron probe. Figure S25, A to D, illustrates the 3D atomic structure around the NV center, reconstructed using single-slice ptychography without (A and B) and with (C and D) probe incorporation. Due to the diamond nanocrystal's thickness being approximately three times the focal depth of the electron probe, pAET reconstructions based on single-slice ptychography were unable to resolve individual atoms. However, multislice ptychography significantly improves the pAET results (*24*, *25*). Figure S25, E and F, presents the 3D atomic structure of the NV center in the xy and xz planes, reconstructed using multislice pAET without probe integration. These results appear less defined compared to the reconstruction with probe integration (fig. S25, G and H), where carbon atoms and the vacancy are more distinctly resolved (fig. S25I). Quantitatively, multislice pAET with probe integration provides the highest precision, achieving an RMSD of 7 pm—2 pm better than reconstructions without probe integration (fig. S25J). Despite this high precision, distinguishing nitrogen atoms from carbon atoms in diamond using pAET remains challenging, as it requires a higher electron dose due to the minimal atomic number difference between nitrogen and carbon.

Multislice pAET simulations of an amorphous carbon nanoparticle

A 3D atomic model of an amorphous carbon nanoparticle, 5 nm in diameter and containing 7,514 carbon atoms, was constructed. The structure was generated in LAMMPS using the Tersoff potential via a melt-quenching process (*74*), with a quenching rate of 0.3 K/ps to ensure high-quality amorphous configurations. The resulting 3D atomic coordinates were used as input for multislice ptychographic simulations performed with abTEM (*60*). Simulation parameters included a 70 kV accelerating voltage, a probe with 13 nm underfocus, a convergence semi-angle of 32 mrad, a collection angle of 65 mrad, 128 × 128 pixels per diffraction pattern, and a scan step size of 0.652 Å (table S1). Eight frozen phonons were used to account for thermal diffuse scattering. Poisson noise was added to the simulated diffraction patterns to represent a finite electron dose of $1.5 \times 10^4$ $e^-$ $Å^{-2}$ per projection. Given the beam sensitivity of amorphous carbon, a wide angular range with relatively large tilt increments was employed to balance reconstruction quality and radiation damage. Specifically, 37 tilt angles were simulated along a single axis, spanning from -72° to +72° in 4° steps, to generate the ptychographic dataset.

Multislice ptychographic reconstructions were carried out using a maximum likelihood algorithm implemented in fold-slice, a MATLAB package adapted from Ptychoshelves (*71, 72*). A representative reconstructed projection is shown in fig. S26A. The reconstruction framework includes a depth-resolution constraint governed by the combined effects of the probe's semi-convergence angle and the detector's maximum collection angle (*20*, *73*). The real-space pixel size was 0.34 Å. Each reconstruction used a single probe mode and ran for 200 iterations, with step sizes of 0.02 applied to both the object and probe updates. The object volume was discretized into 10 slices of 1 nm thickness, and a regularization factor of 0.5 was applied during reconstruction.

After multislice ptychographic reconstruction at each angle, we applied tomographic reconstruction, atom tracing, and position refinement, yielding a 3D atomic model of the amorphous nanoparticle (fig. S26, B and C). Simulation and reconstruction parameters are summarized in Table S1. Comparison with the ground-truth model yielded RMSDs of 4, 5, 9, and 11 pm along the x, y, z axes, and overall, respectively, with 99% of atoms correctly identified (fig. S26, D to G). The medium-range order rings identified in the refined atomic model closely match those in the ground-truth structure (fig. S26, H and I), validating the method's reliability in disordered systems. These results demonstrate that the pAET workflow developed in this study can achieve high-precision 3D atomic structure determination of amorphous carbon at experimentally relevant conditions.

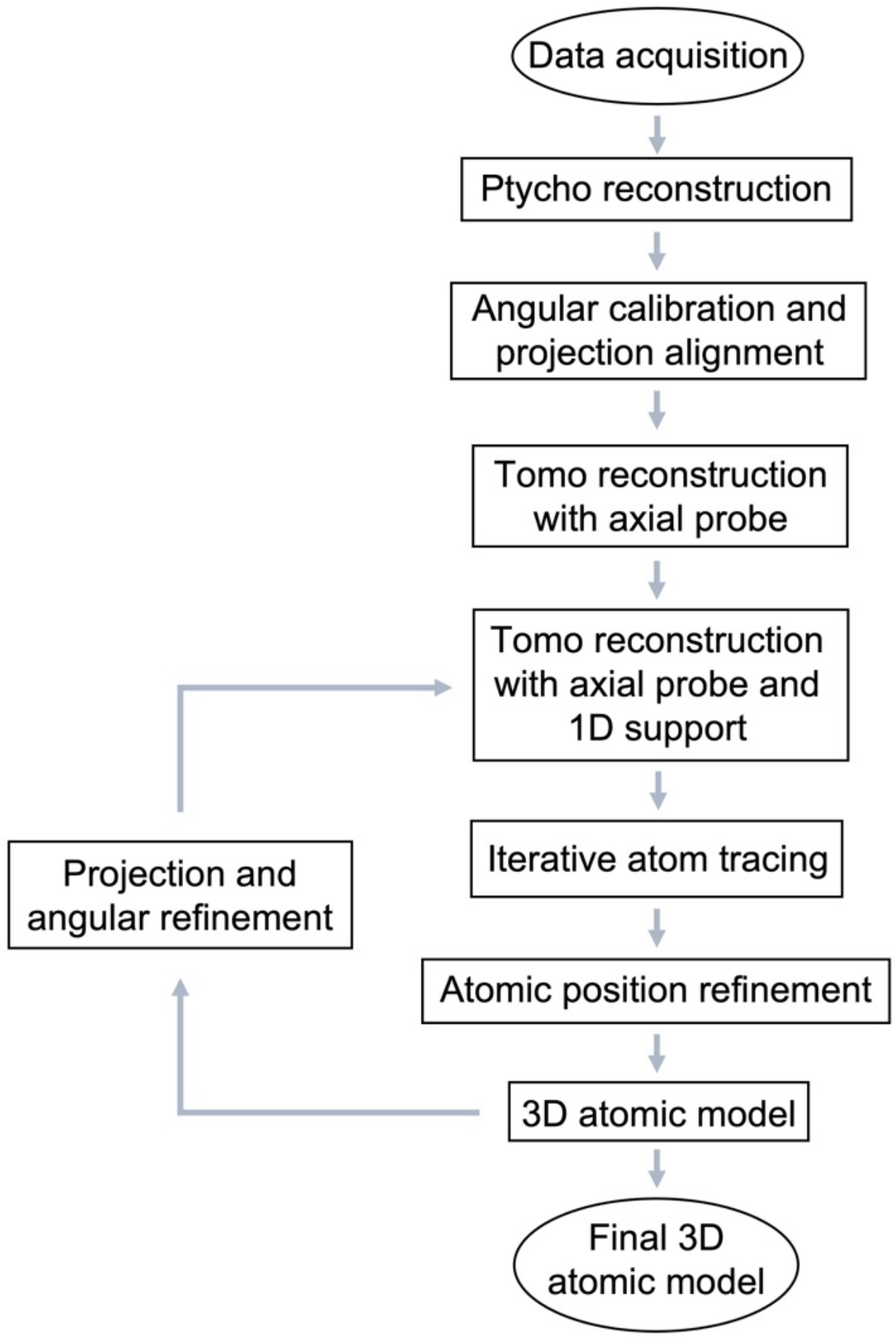

**Fig. S1. Flowchart of the pAET pipeline**. The workflow begins with data acquisition, followed by ptychographic reconstruction of the diffraction patterns to recover phase images. These phase projections are aligned through angular calibration and projection alignment. An initial tomographic reconstruction with an axial probe kernel retrieves 3D structural information, followed by another reconstruction with an axial probe and 1D z-support to enhance depth resolution. Iterative atom tracing identifies individual atomic positions, while atomic position and angular refinements further improve accuracy and consistency across tilts. The refined parameters are used for the final tomographic reconstruction with the axial probe and 1D z-support, followed by iterative atom tracing and refinement, yielding the final 3D atomic model.

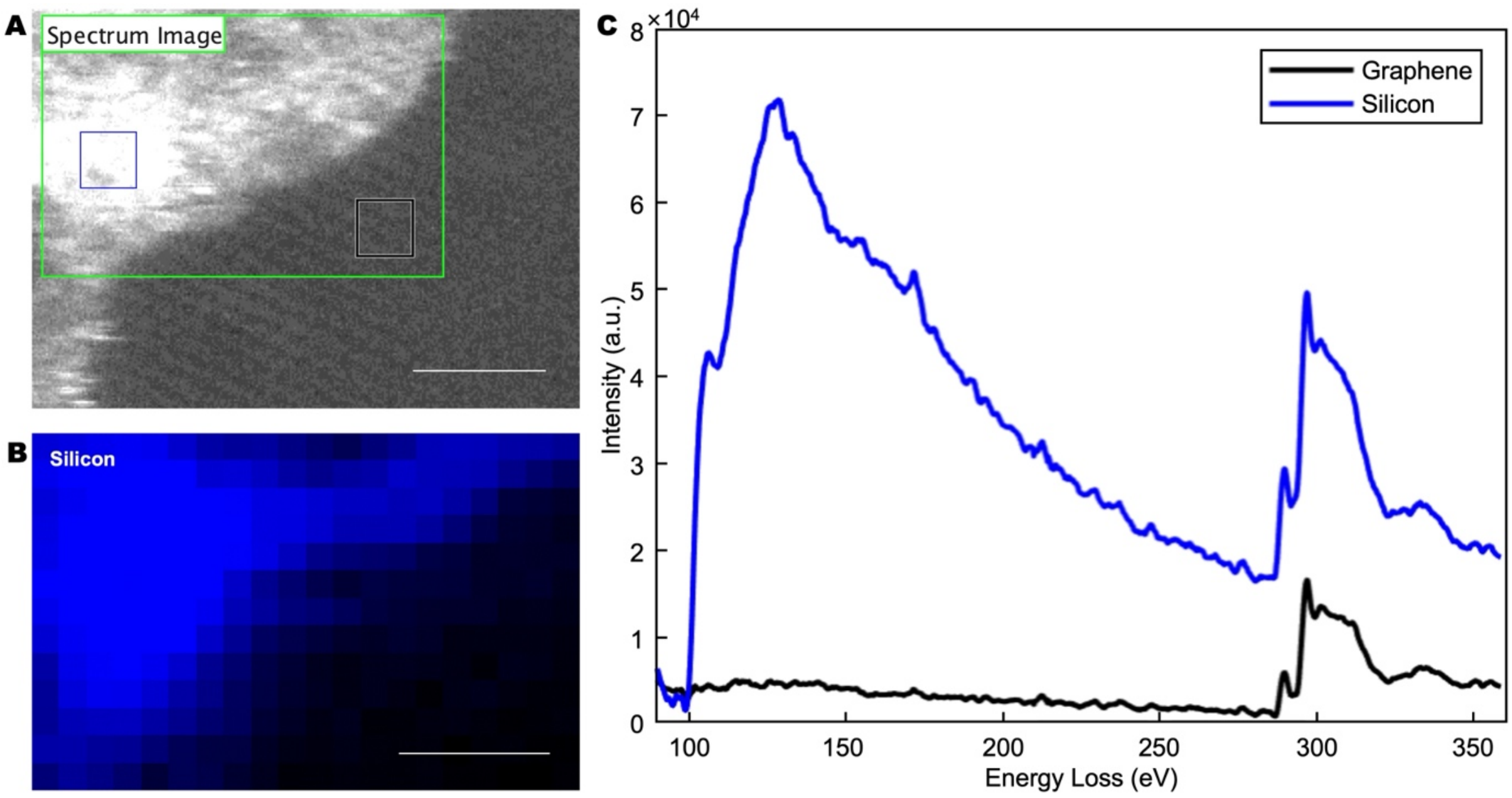

**Fig. S2. Creation of individual Si dopants in TBG.** (**A**, **B**) Annular dark-field (ADF)-STEM images and elemental mapping of Si $L_{2,3}$ edges, highlighting a region with an abundance of dopant resources. (**C**) Electron energy loss spectra of two rectangular insets from (A). Spectral analysis of the resource-rich region (blue) reveals a higher concentration of Si compared to the TBG region (black), confirming selective Si distribution. An electron probe was employed to transfer individual Si atoms from the resource-rich area into defective sites within TBG, facilitating precise dopant incorporation at the atomic scale. Scale bars, 1 nm.

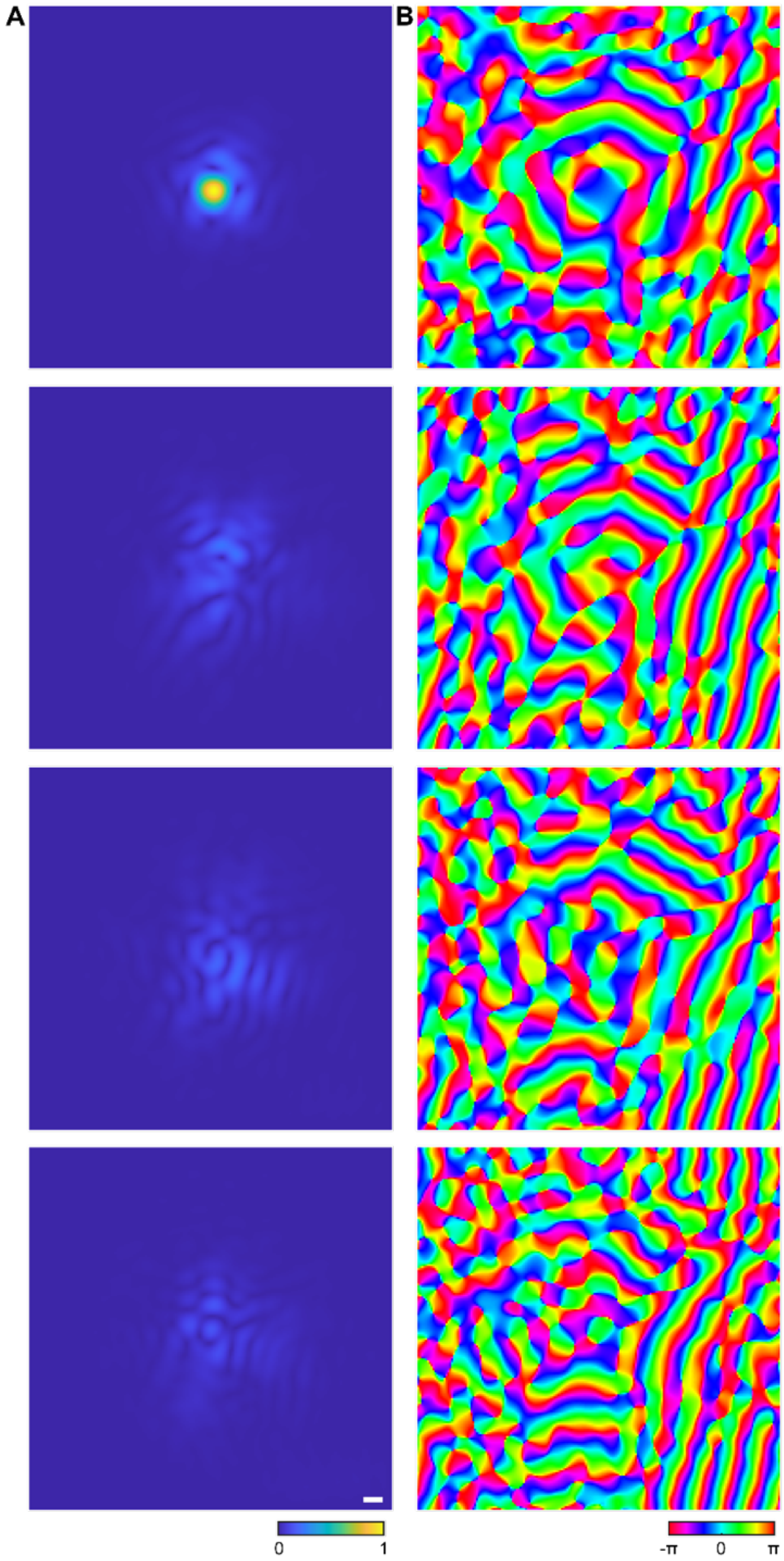

**Fig. S3. Four orthogonal modes of the electron probe. (A, B)** Magnitude and phase distributions of the four orthogonal probe modes obtained from the ptychographic reconstruction of the experimental data. These modes account for partial coherence effects, improving the accuracy of the reconstructed phase. The compositional fractions of the four modes are 72.6%, 13.1%, 9.5%, and 4.8%. Scale bar, 1 Å.

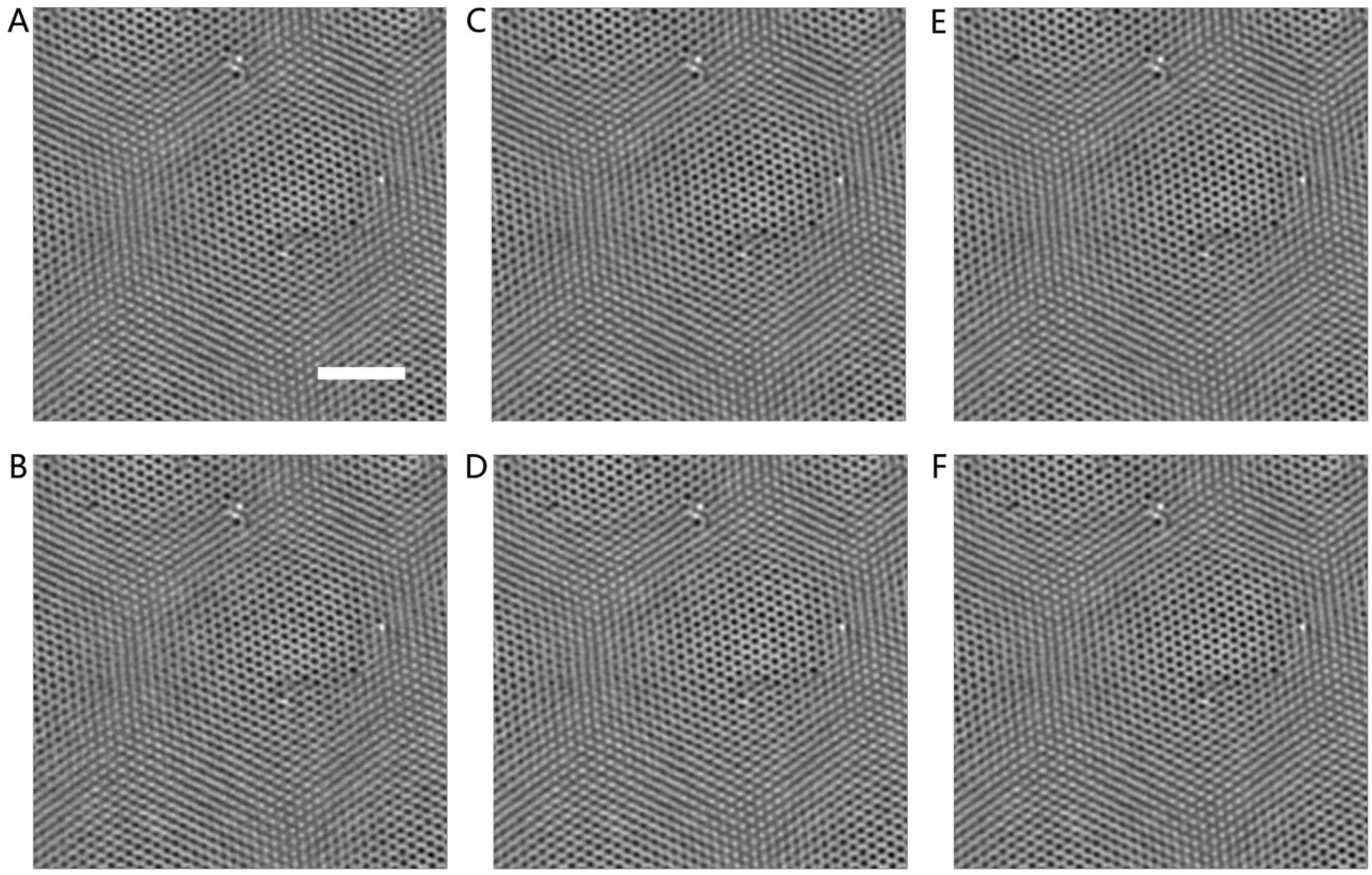

**Fig. S4. Robustness of the ptychographic reconstructions with respect to partial-coherence modeling and detector response**. (**A**, **B**) Reconstructions of a representative experimental dataset using 2 partial-coherence modes without (A) and with (B) a detector response model. (**C**, **D**) Same as (A, B) but with 4 partial-coherence modes. (**E**, **F**) Same as (A, B) but with 6 partial-coherence modes. Scale bar, 2 nm.

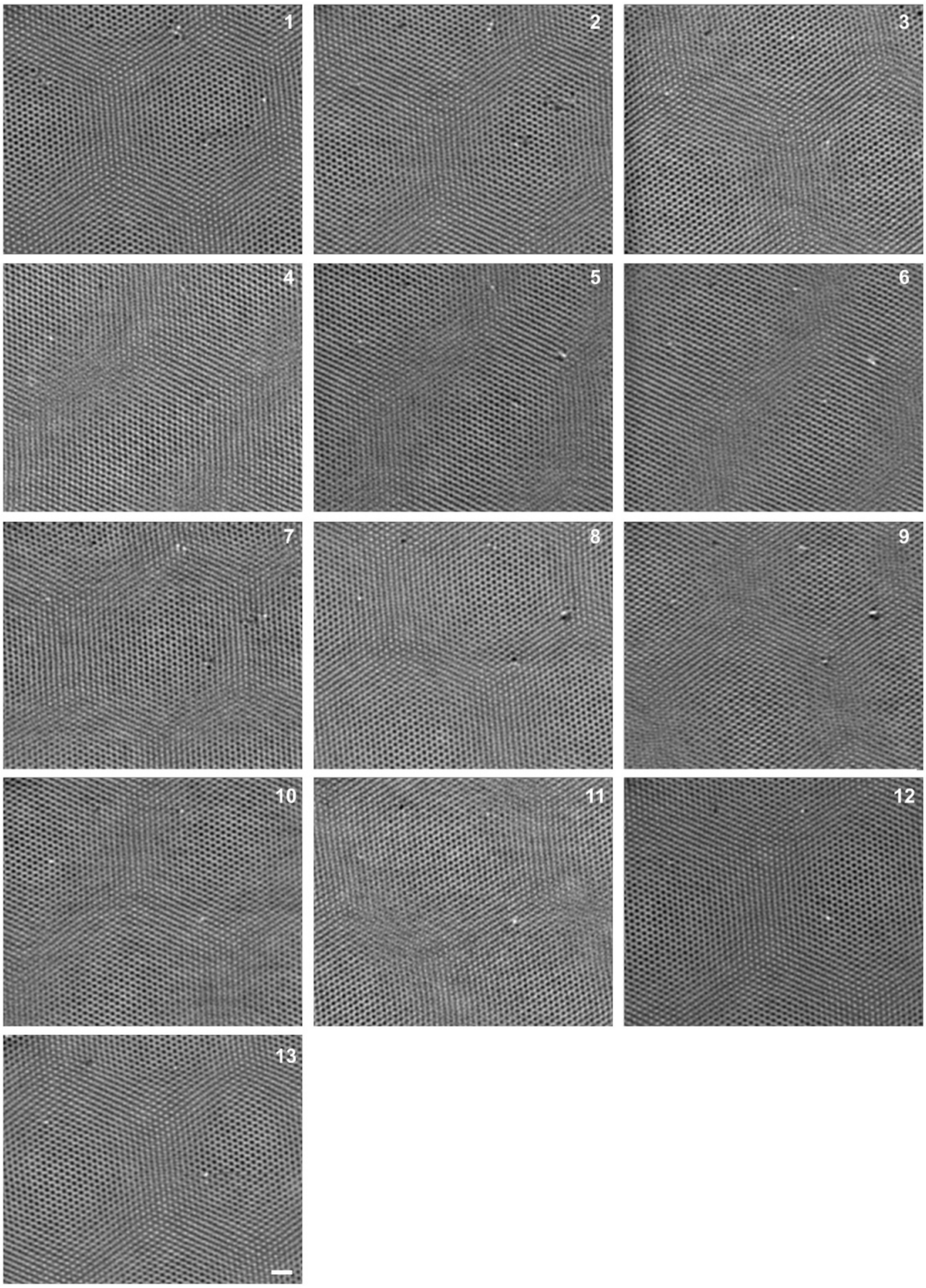

**Fig. S5. Ptychographic phase projections of the experimental TBG data.** Each of the 13 ptychographic reconstructions was obtained from 52,900 diffraction patterns, with bright dots corresponding to individual Si dopants. Careful examination of these dopants shows that their positions remain stable across all projections. Scale bar, 2 nm.

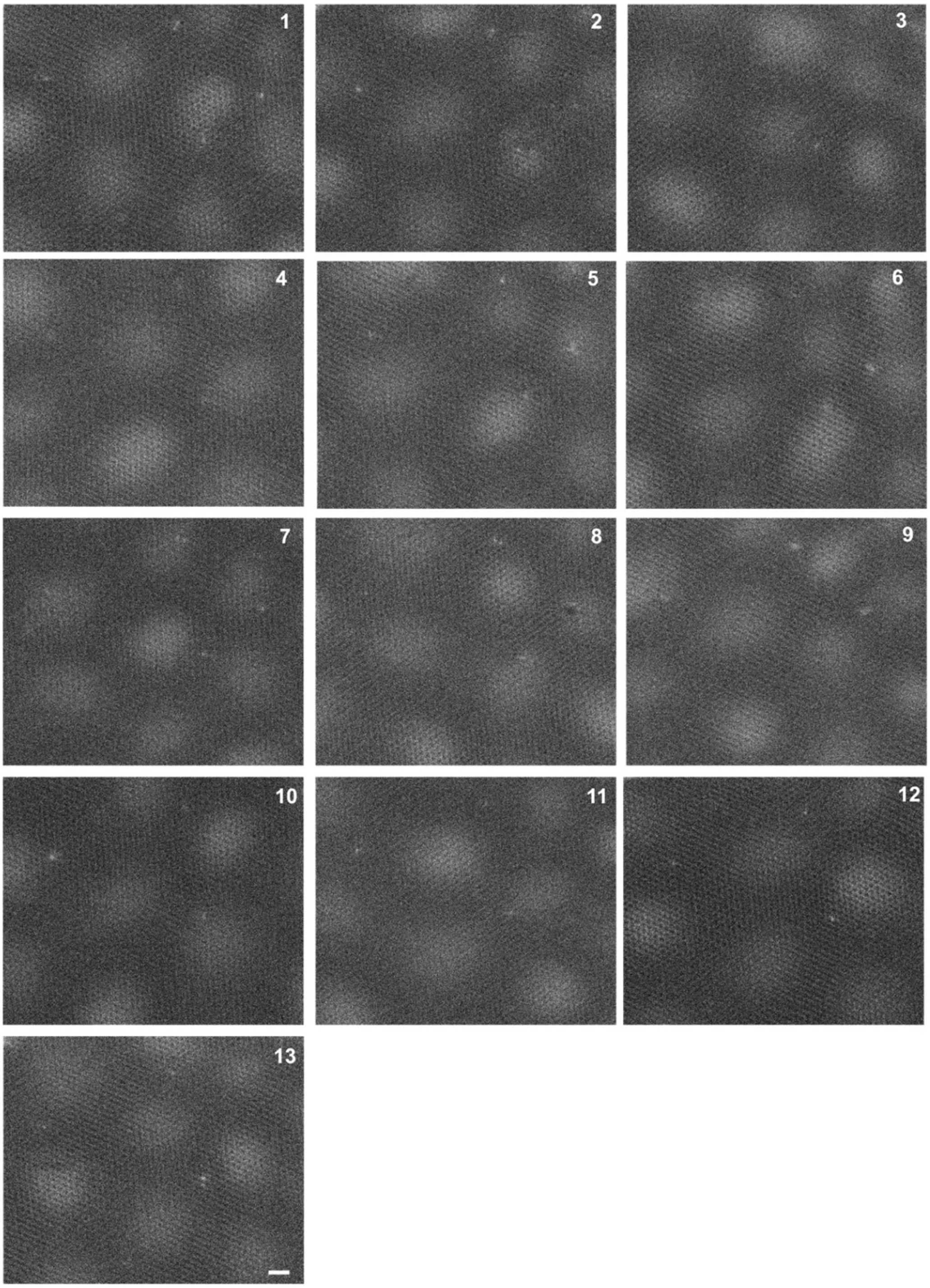

**Fig. S6. Annular dark-field STEM images corresponding to the same regions shown in fig. S5.** Bright contrast marks individual Si substitutional dopants. Scale bar, 2 nm.

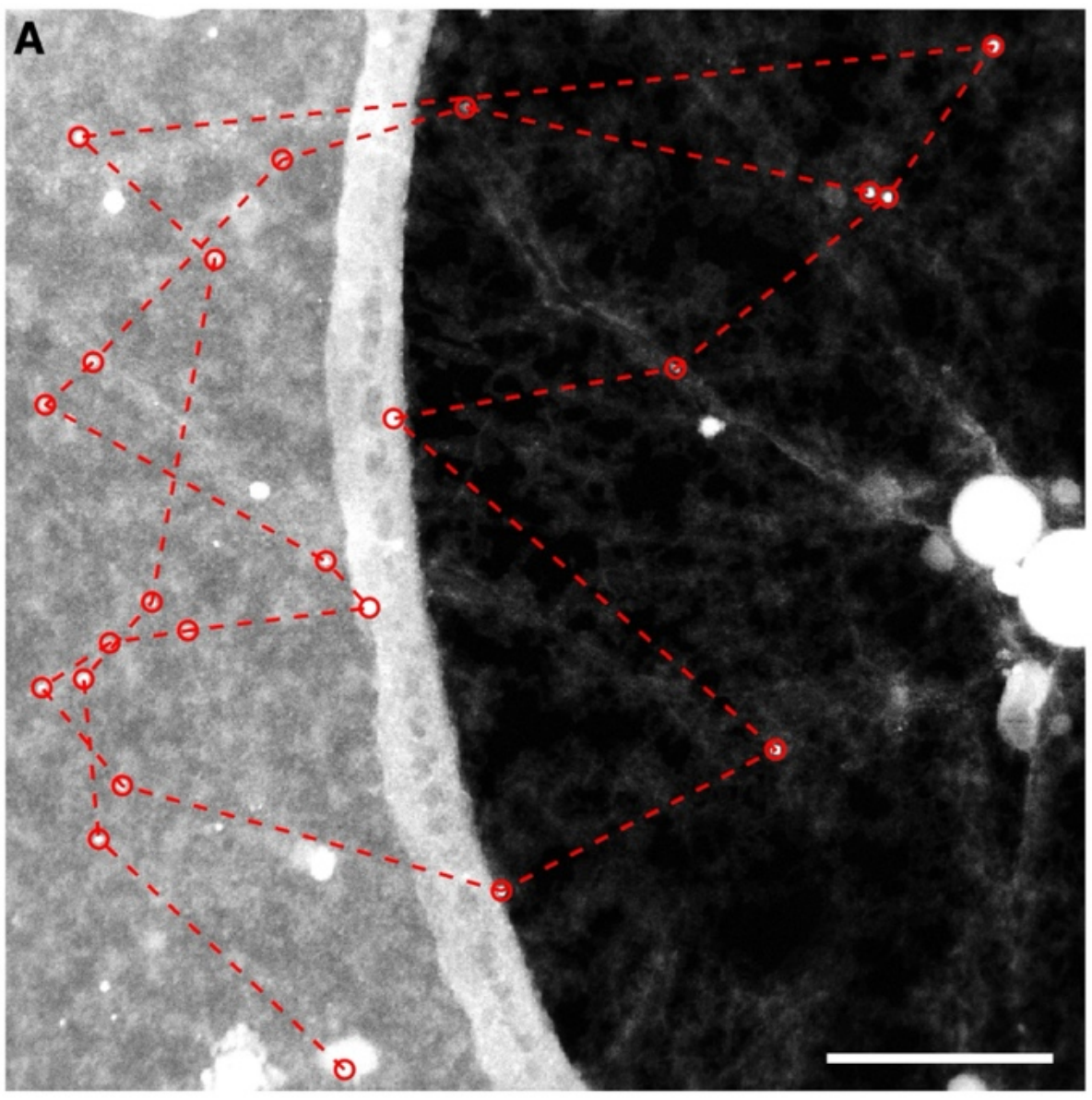

| | Calibrated angles (°) | | | Refined angles (°) | | |
|---|---|---|---|---|---|---|
| Tilt axis | ϕ | θ | ψ | ϕ | θ | ψ |
| Projection # 1 | 0 | 6 | -0.2 | 0 | 7.7 | -0.6 |
| 2 | -0.5 | 3.6 | -8.4 | -0.4 | 4.3 | -7.2 |
| 3 | -1.1 | -9.8 | -11.2 | -1.4 | -7.7 | -9.8 |
| 4 | 0.4 | 15.9 | -11.6 | 0.6 | 15.4 | -11.5 |
| 5 | -0.5 | 9.3 | -14.1 | -0.4 | 9.2 | -13.2 |
| 6 | -1.1 | 5.2 | -17.6 | -1.1 | 3.2 | -16.4 |
| 7 | 0.7 | 14.8 | -4.6 | 0.8 | 16.1 | -3.7 |
| 8 | 0.4 | 9.7 | 8.7 | 0.4 | 8 | 9.1 |
| 9 | -0.4 | 3.6 | 15 | -0.4 | 3.7 | 15.3 |
| 10 | -0.5 | 0.2 | 12.5 | -0.2 | 5.7 | 10.2 |
| 11 | 1.2 | 16 | 10.8 | 1.5 | 19.6 | 10.8 |
| 12 | -0.7 | -5.8 | -0.8 | -0.5 | -6.8 | -0.8 |
| 13 | -0.7 | -5 | -2.7 | -0.7 | -5.7 | -4.5 |

**Fig. S7. Calibration and refinement of the 13 experimental tilt angles.** (**A**) Representative low-magnification annular dark-field image showing nanoparticles (brightest dots). A tilt series of 13 low-magnification images was used to determine the 2D coordinates of these nanoparticles for each projection (red circles), from which the three rotation angles — $\phi$, $\theta$, and $\psi$ — about the x-, y-, and z-axes were calibrated. Scale bar, 100 nm. (**B**) The angles were further refined using the tomographic reconstruction algorithm. The table displays the calibrated and refined three angles for each of the 13 projections.

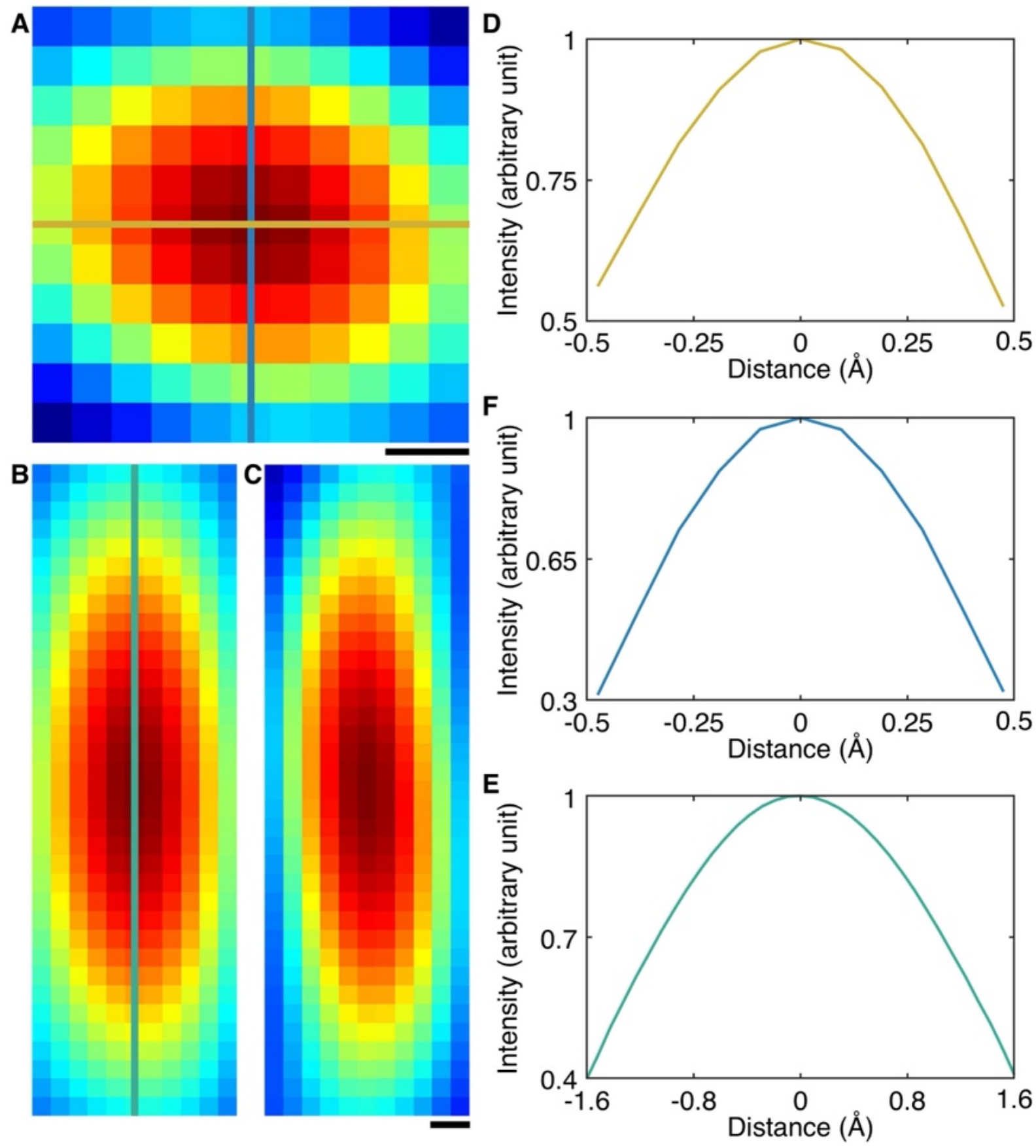

**Fig. S8. Quantification of 3D resolution in pAET using the experimental TBG dataset without a 1D z-support.** (**A–C**) Orthogonal slices in the xy, xz, and yz planes from a 3D volume obtained by statistically averaging 862 representative carbon atoms reconstructed without a z-axis support constraint. The z axis corresponds to the missing-wedge direction. (**D–F**) 1D intensity profiles extracted through the centers of the slices in (A–C). Fitting with a 3D Gaussian function yields resolutions of 1.01, 0.73, and 2.94 Å along the x, y, and z axes, respectively. The lower resolution along the z axis arises from the missing-cone problem due to the limited tilt range of the double-tilt holder, while the slight difference between x- and y-axis resolutions is attributed to asymmetry in the angular coverage of the 13 projections. Scale bars, 0.2 Å.

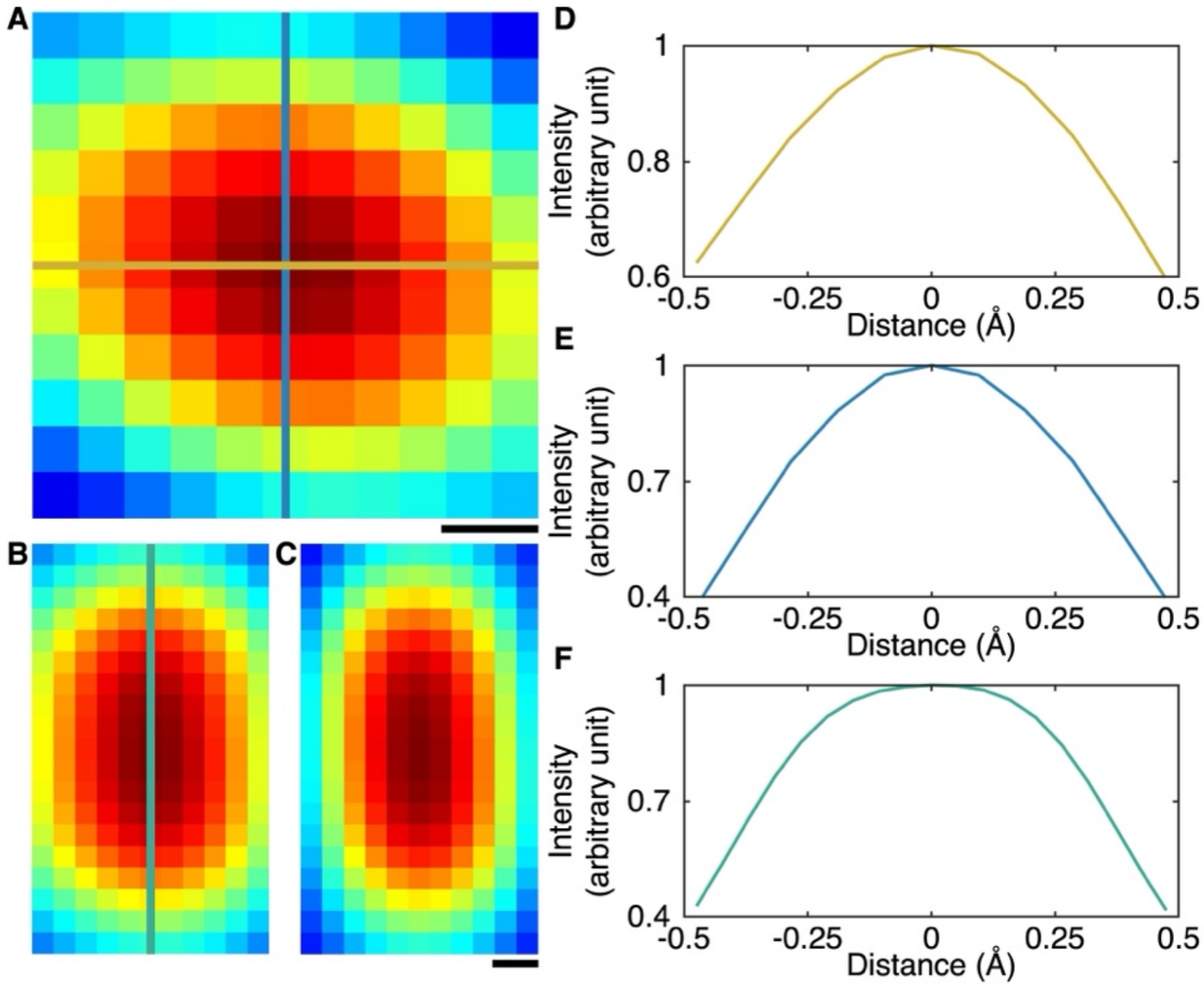

**Fig. S9. Quantification of 3D resolution in pAET using the experimental TBG dataset with a 1D z-support.** (**A**–**C**) Orthogonal slices in the xy, xz, and yz planes from a 3D volume obtained by statistically averaging 862 representative carbon atoms reconstructed with a z-axis support constraint. The z axis corresponds to the missing-wedge direction. (D–F) 1D intensity profiles extracted through the centers of the slices in (A–C). Fitting with a 3D Gaussian function yields resolutions of 1.09, 0.76, and 1.49 Å along the x, y, and z axes, respectively. Scale bars, 0.2 Å.

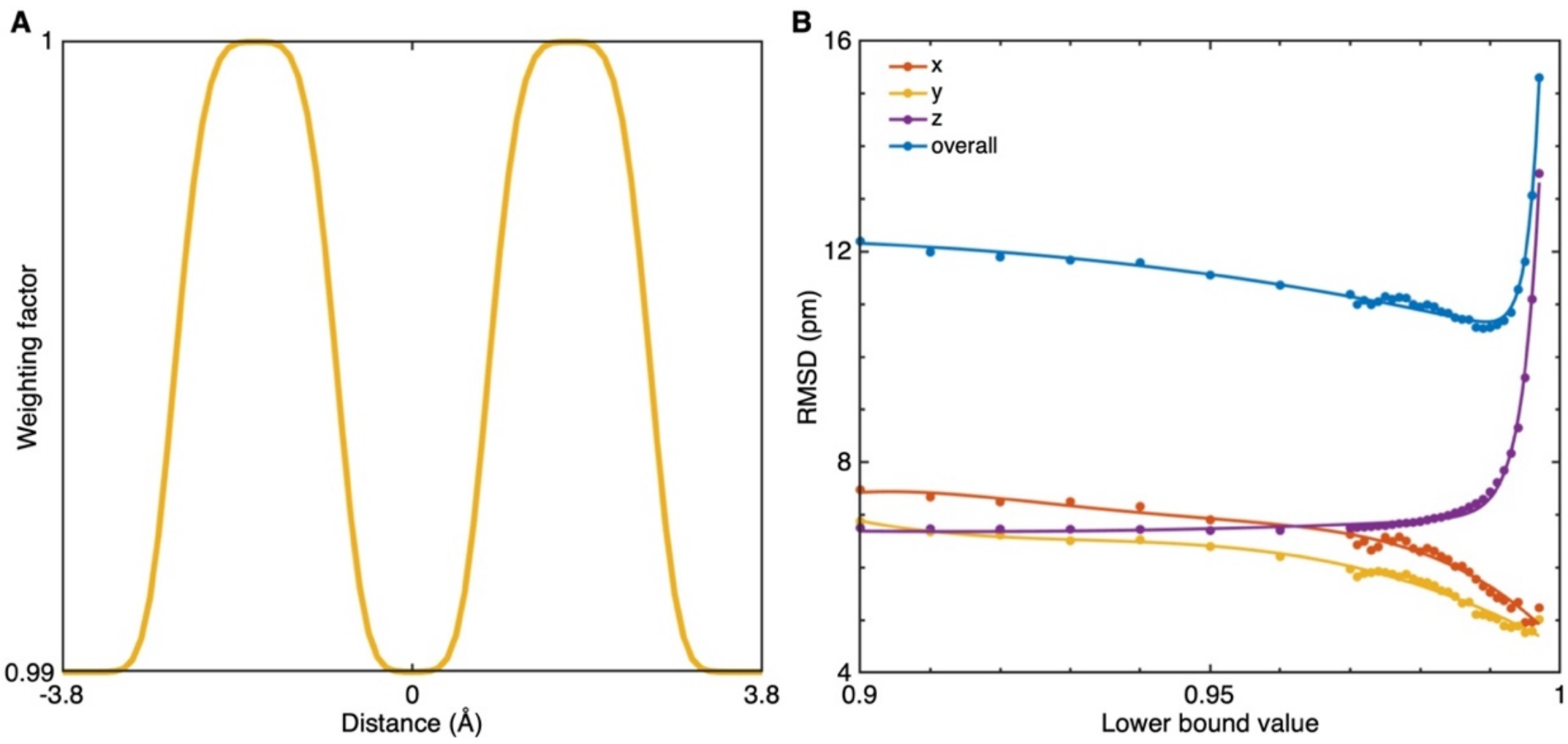

**Fig. S10. Optimization of the 1D support along the z axis.** (**A**) A pair of generalized Gaussian functions was used as a 1D support constraint during tomographic reconstruction. The function was normalized between a lower bound value (0.99 in this case) and 1 and applied multiplicatively along the z axis in each iteration. (**B**) RMSDs along x, y, z, and overall as a function of the lower bound value. Tighter support corresponds to smaller lower bound values. Each data point represents an independent 3D reconstruction and atom tracing, and the solid lines are fits to the data. The optimal value is 0.99, yielding an overall RMSD of 11 pm. As the lower bound approaches 1, the overall RMSD increases sharply because the absence of an effective support constraint causes overlapping carbon atoms along the z axis, reducing the accuracy of the automated tracing algorithm and leading to larger z-coordinate errors.

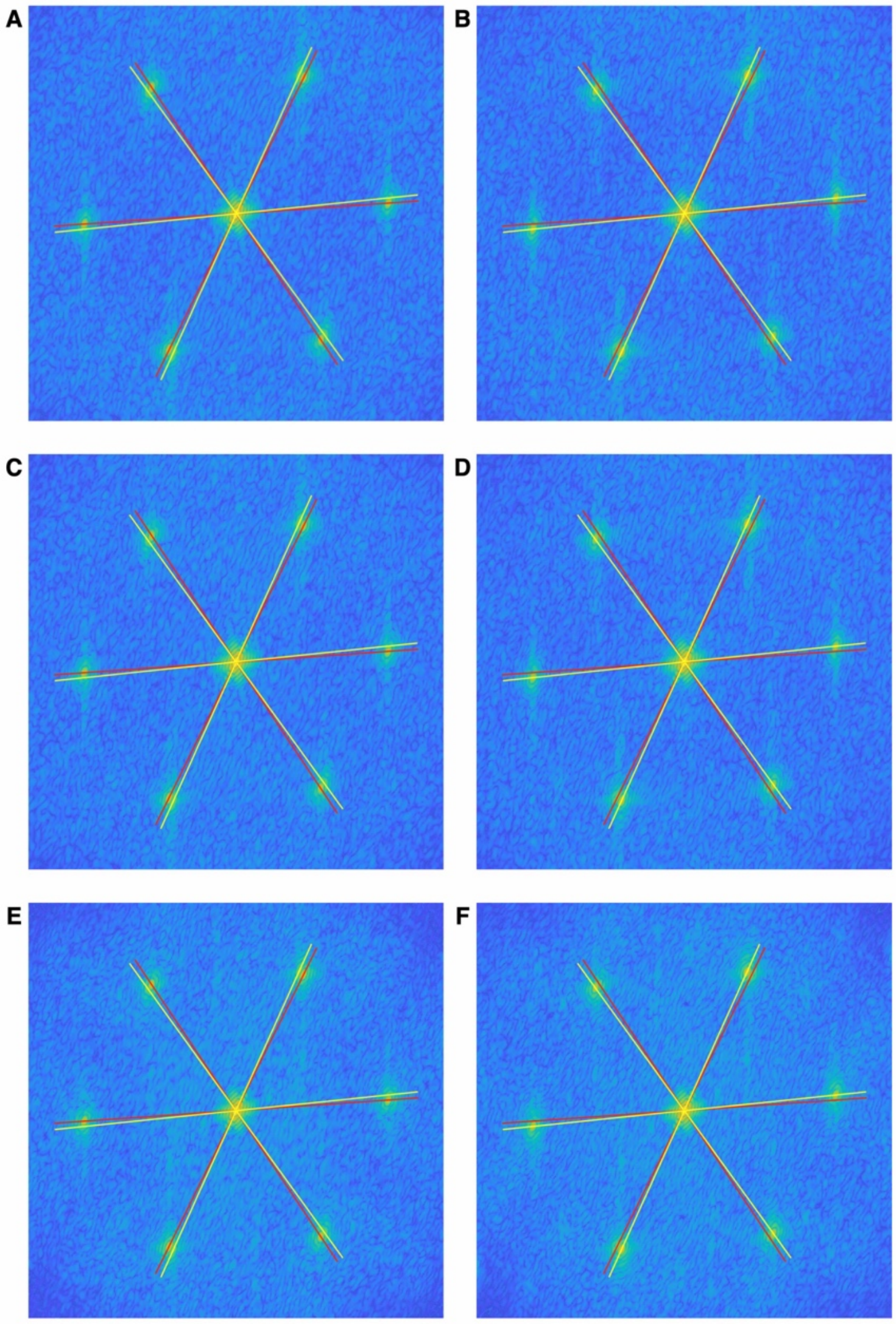

**Fig. S11. Separation of top and bottom layers in experimental and simulated TBG.** (**A**–**D**) Fourier transforms of 5-voxel-thick slabs (0.48 Å) extracted from the top (left) and bottom (right) layers of the experimental 3D reconstruction, obtained without (A, B) and with (C, D) the 1D z-support constraint. (**E**, **F**) Corresponding Fourier transforms from the 3D reconstruction based on multislice-simulated projections with the 1D z-support, which are highly consistent with those obtained without the support. Red and yellow lines indicate Bragg peak positions from the top and bottom layers, respectively, demonstrating clear layer separation in both experimental and simulated data.

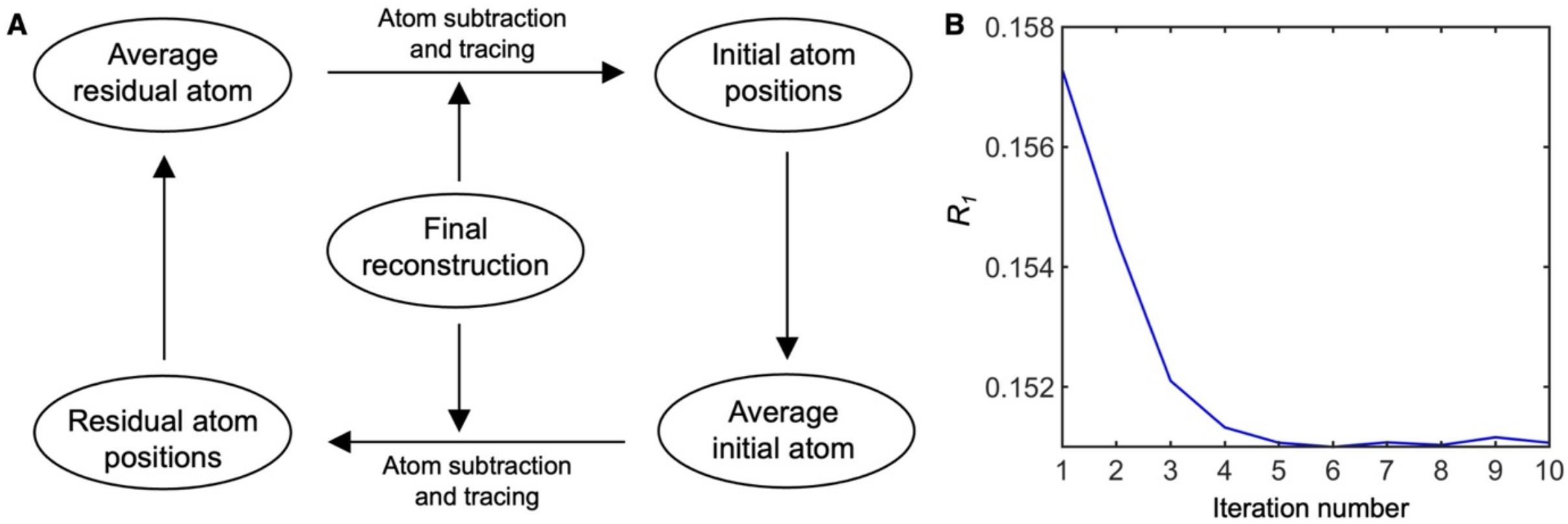

**Fig. S12. Diagram of an iterative atom-tracing method.** (**A**) The method started by tracing atoms from the final reconstruction using the polynomial method. After determining a set of 3D coordinates for the initial atoms, an average initial atom was calculated. The final reconstruction was then subtracted by the average initial atom at each traced position, and the polynomial method was applied again to trace the residual atoms, from which an average residual atom was generated. This average residual atom, along with the traced positions, was used in the next iteration (*37*). (**B**) Decrease in the $R_1$ factor with increasing iteration number.

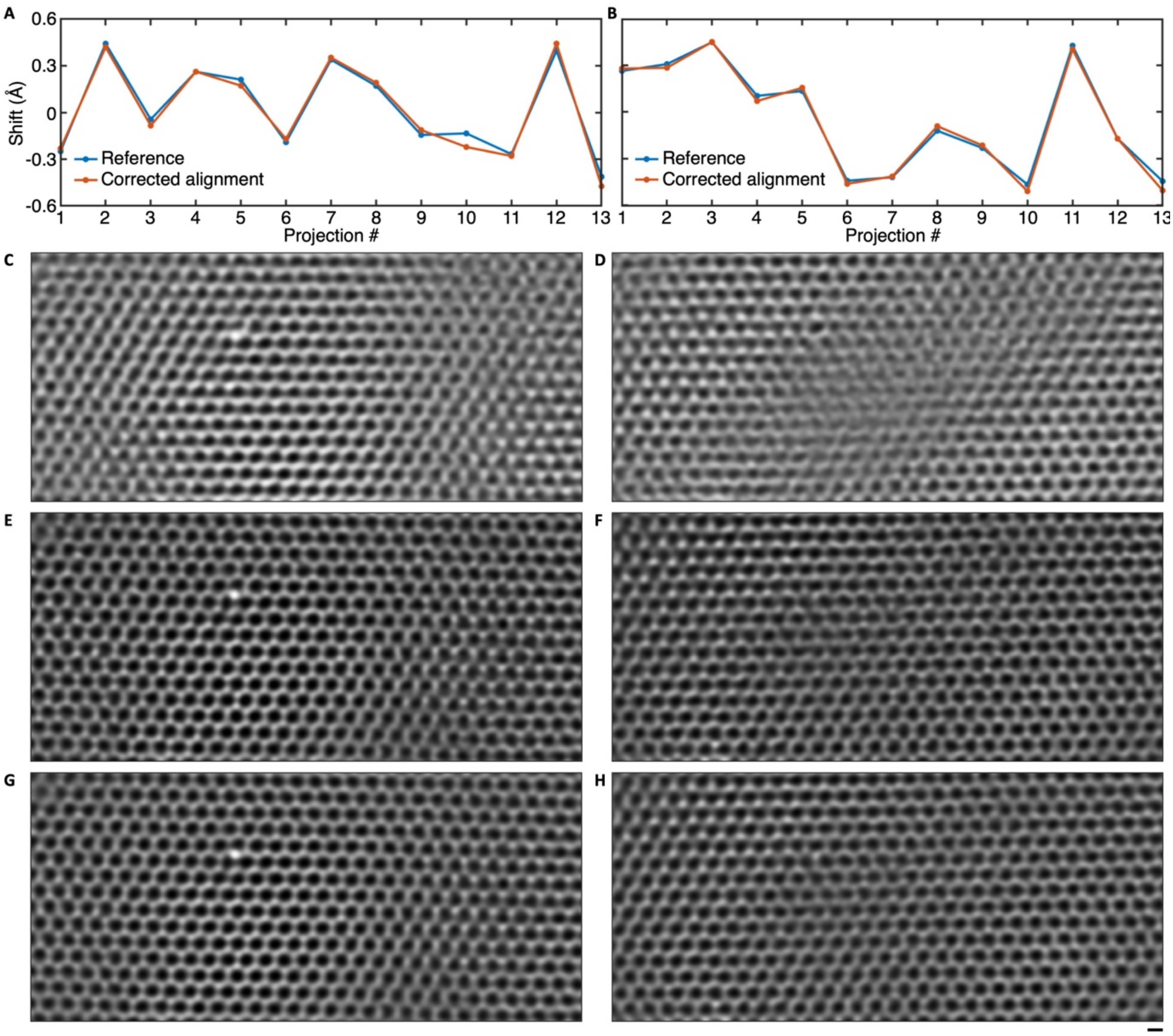

**Fig. S13. Robustness of the automated projection alignment during atom refinement.** (**A**, **B**) Random in-plane shifts of –0.5 Å to +0.5 Å were applied along the x (A) and y (B) axes to each of the 13 experimental projections within a selected region of interest (blue = reference). The refinement procedure corrected these misalignments (orange). (**C**, **D**) Reconstructed top (C) and bottom (D) layers before correction. (**E**, **F**) Corresponding layers after correction. (**G**, **H**) Reference reconstructions from the original 13 projections without induced shifts. The cross-correlation coefficients are 90.5% between (C) and (G) and 88.3% between (D) and (H) before correction, improving to 99.6% between (E) and (G) and 99.2% between (F) and (H) after correction. Scale bar, 2 Å.

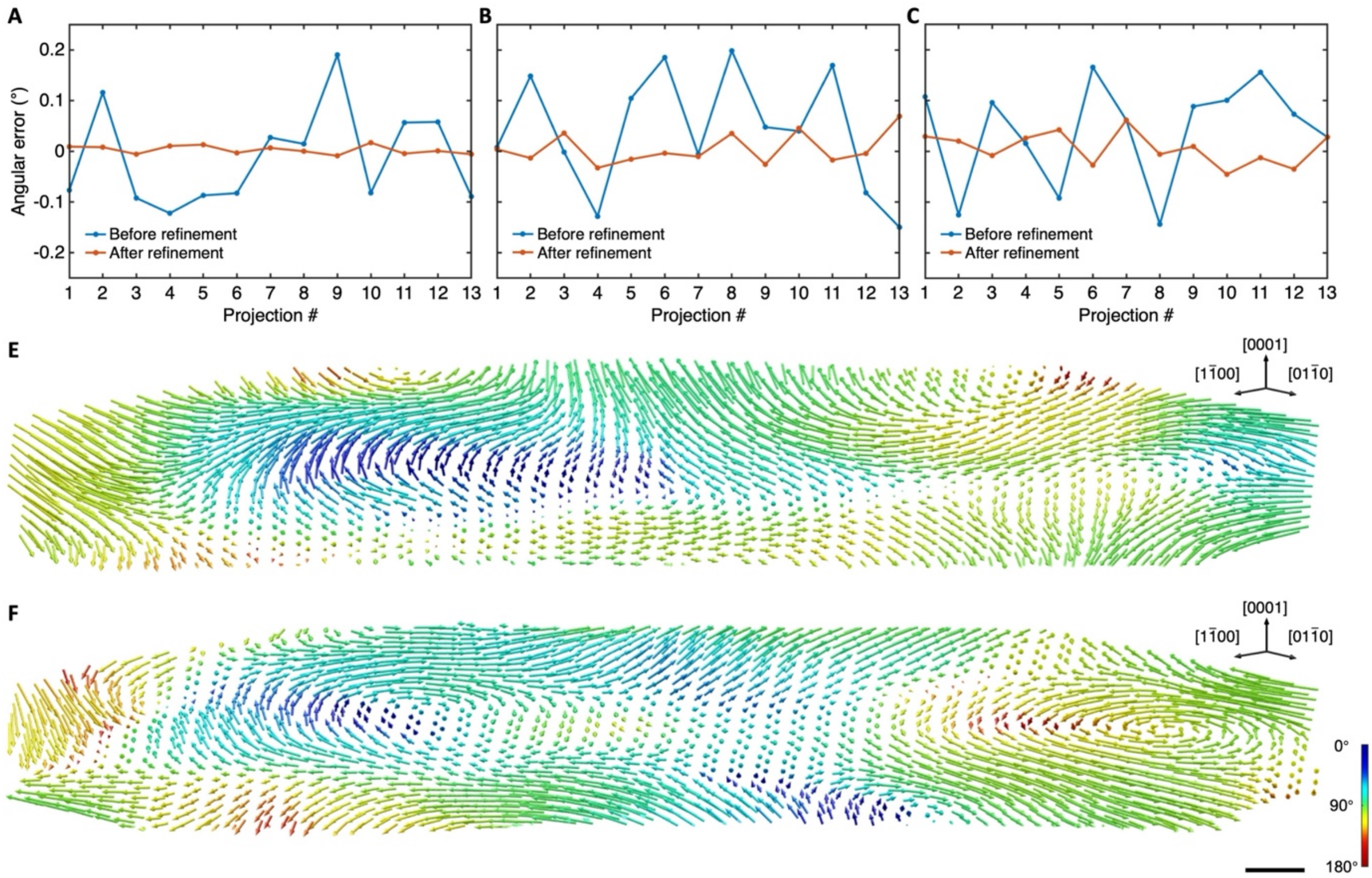

**Fig. S14. Angular refinement and resulting 3D displacement vector fields.** (**A**–**C**) Random shifts between –0.2° and +0.2° were introduced to the three rotation angles — ϕ (A), θ (B), and ψ (C) — about the x-, y-, and z-axes for all 13 projections (blue = reference). The refinement procedure corrected these misalignments (orange). (**E**, **F**) 3D displacement vector fields in the top and bottom layers after refinement, showing good agreement with Fig. 3A and B in the main text. Vector color indicates the polar angle relative to the xy plane. Atomic positions were convolved with a 3D Gaussian kernel, with widths corresponding to the in-plane lattice constant and the average interlayer spacing along z, to enhance statistical precision. Vector magnitudes are scaled by a factor of 25 for visualization. Scale bar, 5 Å.

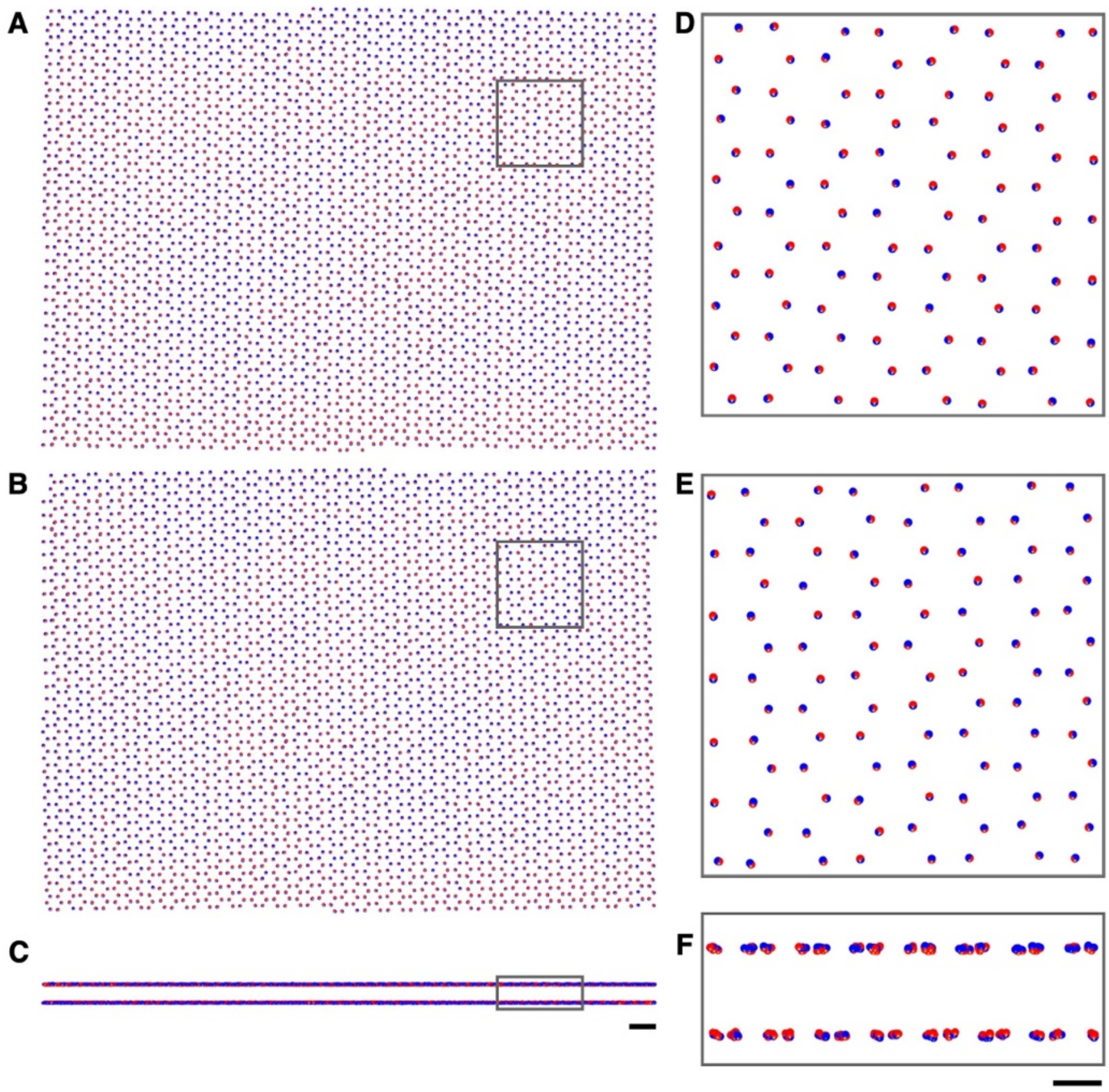

**Fig. S15. Comparison of 3D atomic positions between the randomly perturbed (ground-truth) and pAET-reconstructed models.** (**A, B**) Top and bottom layers of the two models, with red and blue dots indicating carbon atoms in the ground-truth and reconstructed models, respectively. (**C**) Side view of the two models. (**D-F**) Magnified views of the boxed regions in panels (A-C), respectively. Scale bars: 5 Å in (C), 2 Å in (F).

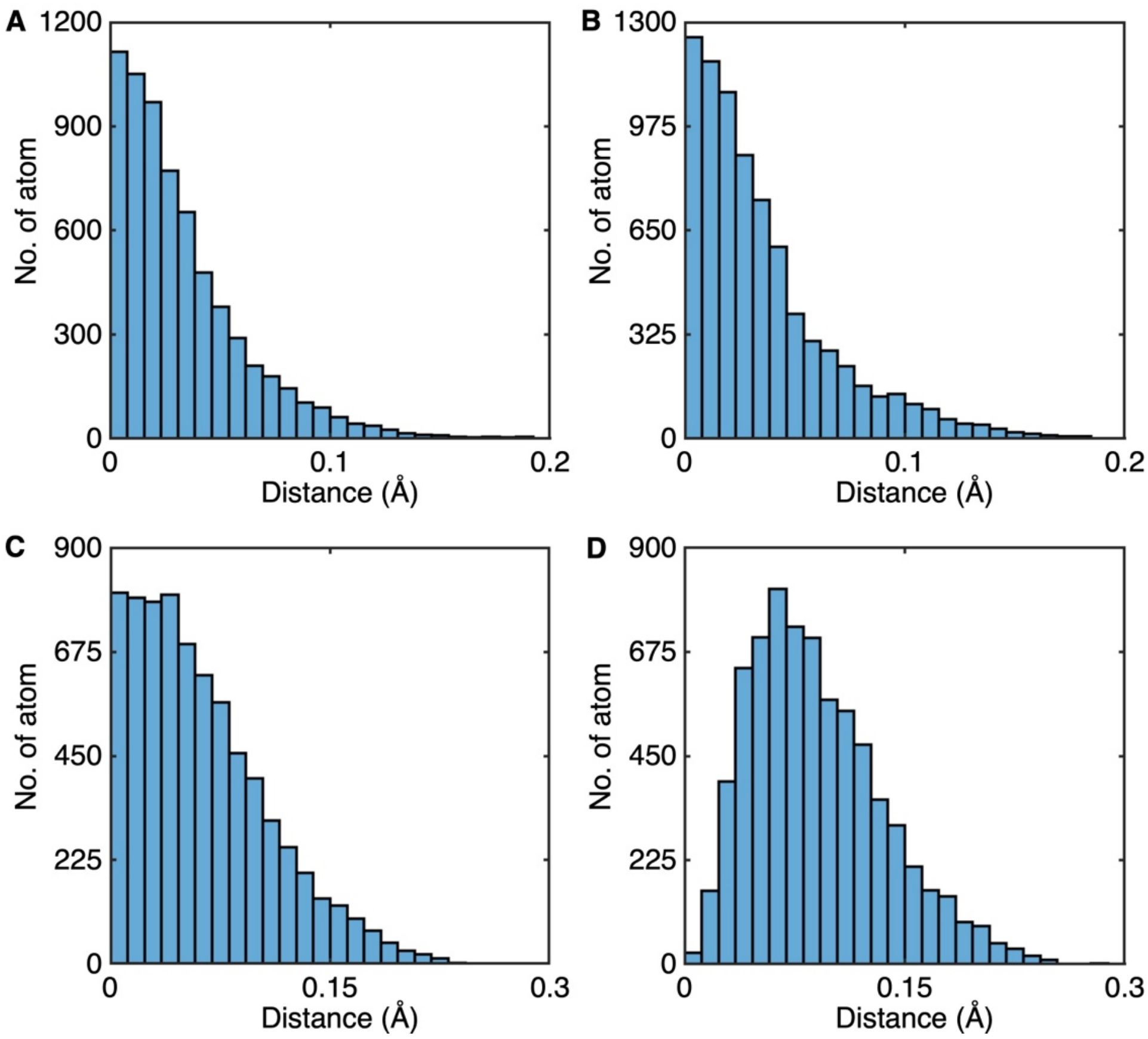

**Fig. S16. Quantification of 3D precision between the randomly perturbed and pAET-reconstructed models.** (**A**-**D**) RMSDs along the x, y, and z axes, and overall, are 4, 5, 8, and 10 pm, respectively.

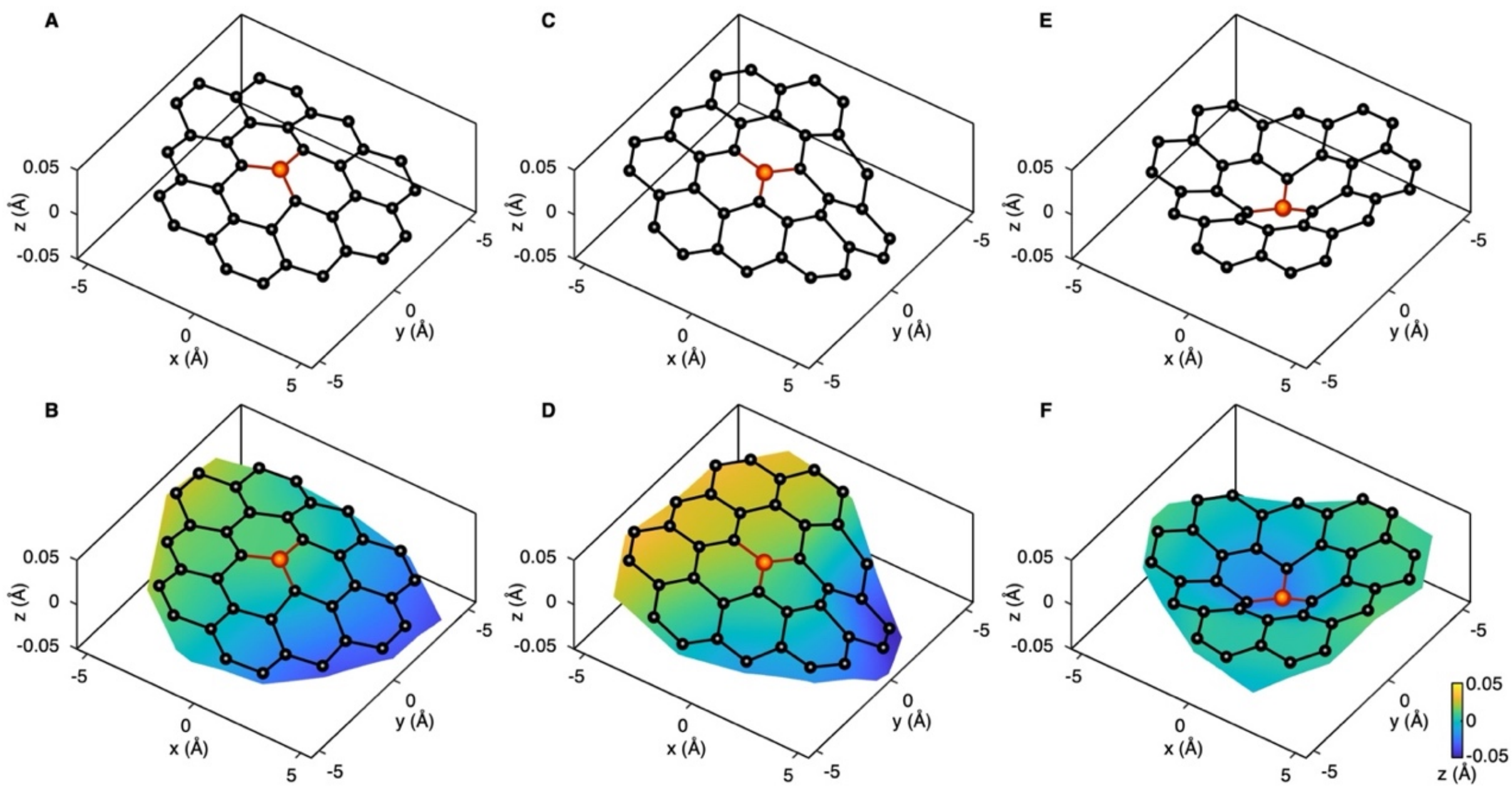

**Fig. S17. 3D visualization of the Si dopants identified in the studied TBG.** (**A**, **B**) Dopant located in the top layer, showing minimal local distortion, with color indicating the z-axis deviation. (**C**, **D**) Dopant located in the bottom layer, also exhibiting minimal local distortion. (**E**, **F**) Another dopant in the bottom layer, displaying a pronounced local lattice distortion.

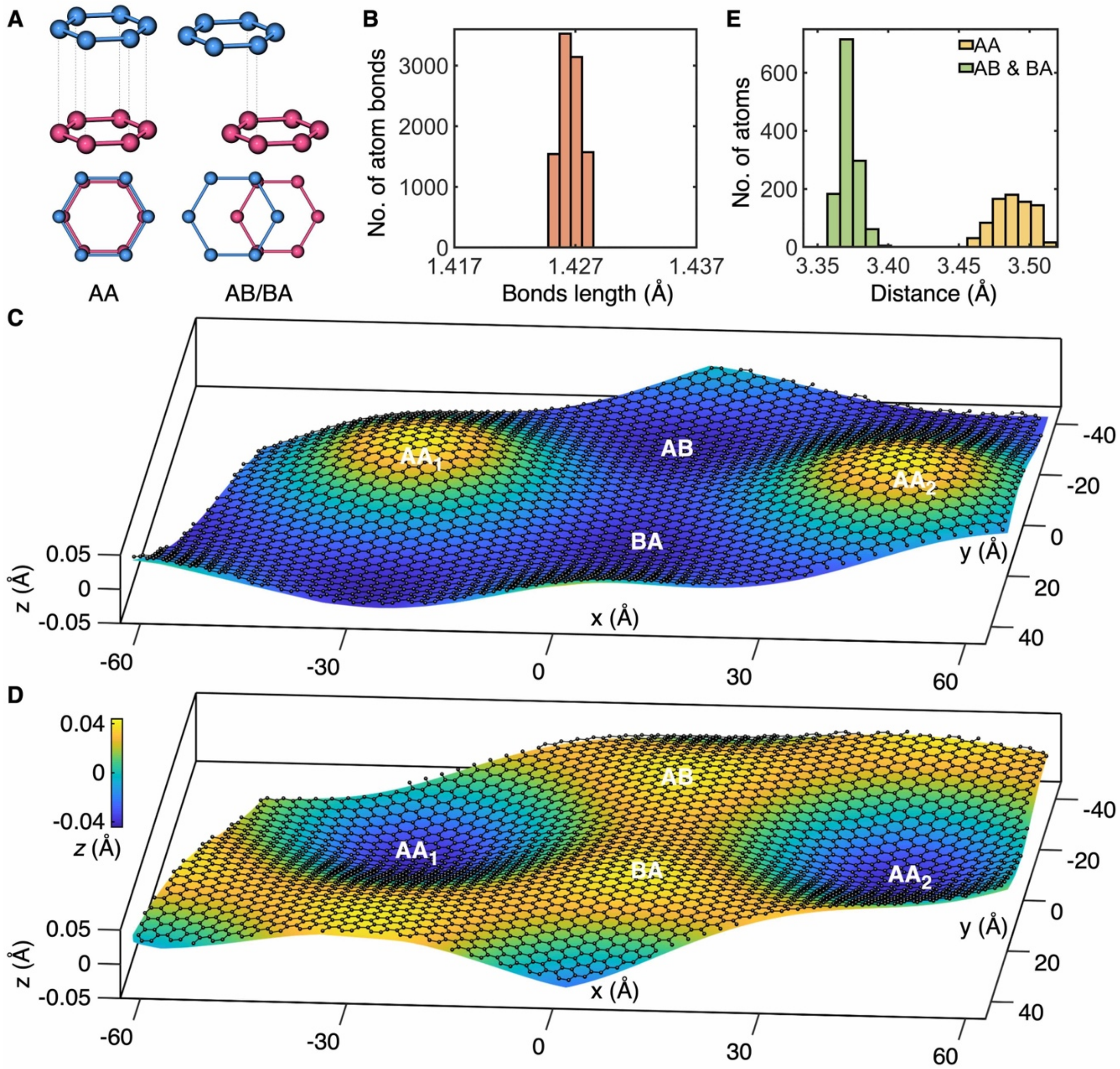

**Fig. S18. 3D atomic structure of the DFT-relaxed TBG model.** (**A**) Schematic of AA and AB/BA stacking in TBG. (**B**) Histogram of $sp^2$ carbon-carbon bond lengths in the DFT-relaxed model, with a mean value of 1.425 Å. (**C**) Histogram of interlayer distances in the AA and AB/BA stacking regions of the relaxed model, showing mean values of 3.49 Å and 3.37 Å, respectively. (**D**, **E**) Variations in the atomic coordinates in the z-direction with respect to a flat model for (D) top and (E) bottom layers of the relaxed atomic model. The yellow regions indicate upward curvature and blue regions indicate downward curvature within the layers.

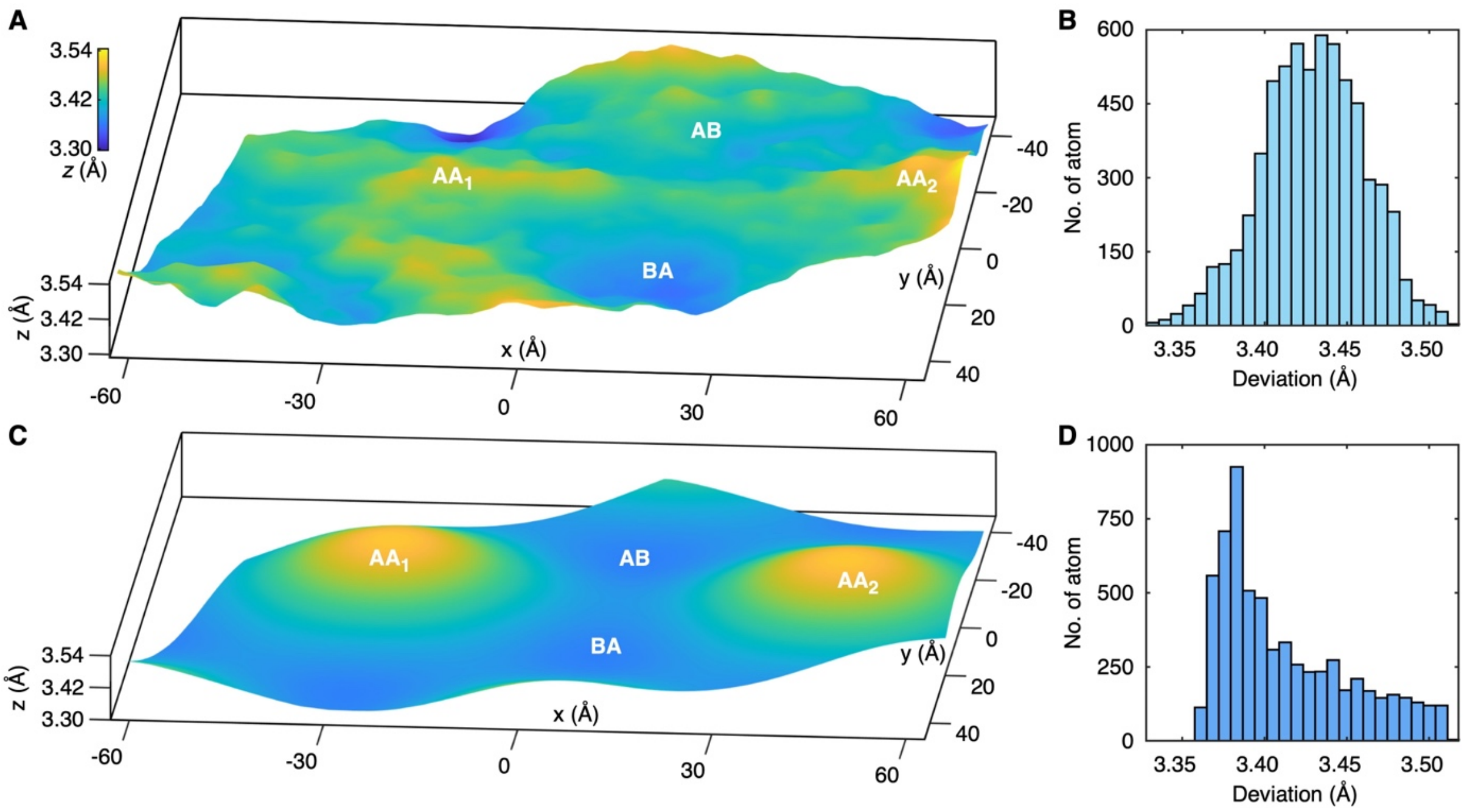

**Fig. S19. Quantitative characterization of the interlayer spacing in TBG.** (**A**) Variations in interlayer spacing for the experimental atomic model. (**B**) Histogram of the interlayer spacing for the experimental atomic model, with a mean of 3.43 Å and a standard deviation of 0.03 Å. (**C**) Variations in interlayer spacing for the relaxed atomic model. (**D**) Histogram of the interlayer spacing for the relaxed atomic model, with a mean of 3.41 Å and a standard deviation of 0.04 Å.

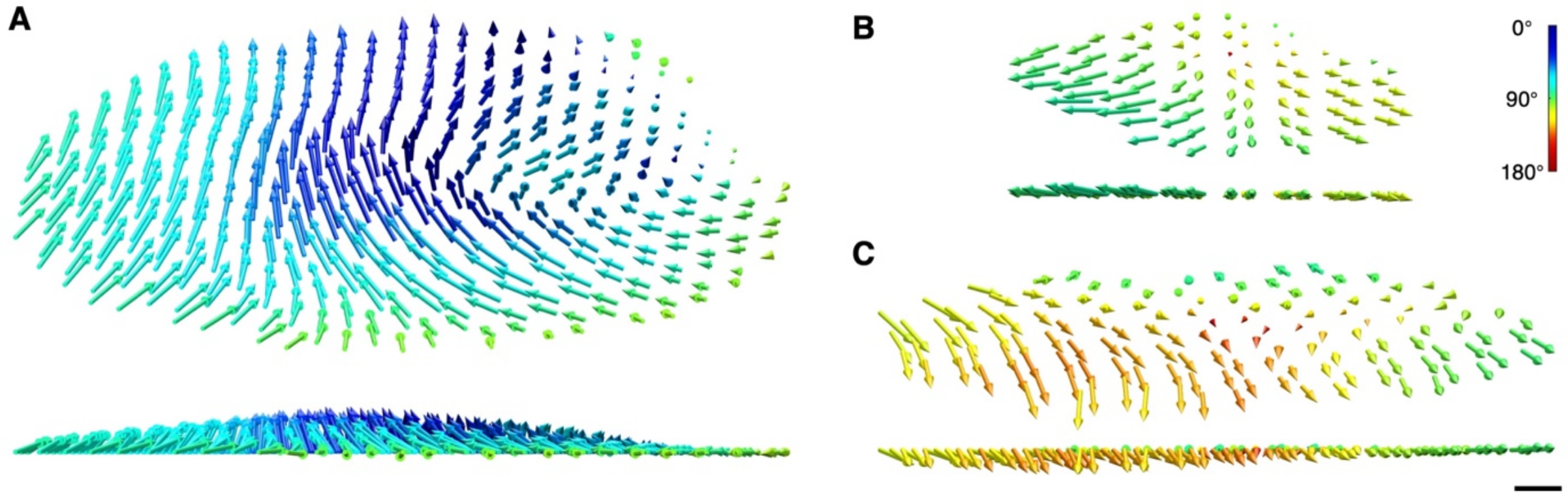

**Fig. S20. 3D displacement vectors in TBG revealed by pAET. (A–C)** 3D vector fields corresponding to two anti-meron-like (A, C) and one meron-like (B) structural textures, with respective skyrmion numbers $N_s$ = 0.35, 0.44, and −0.42. Vector color indicates the polar angle relative to the xy plane. Vector magnitudes are scaled by a factor of 25 for visualization. Scale bar, 2 Å.

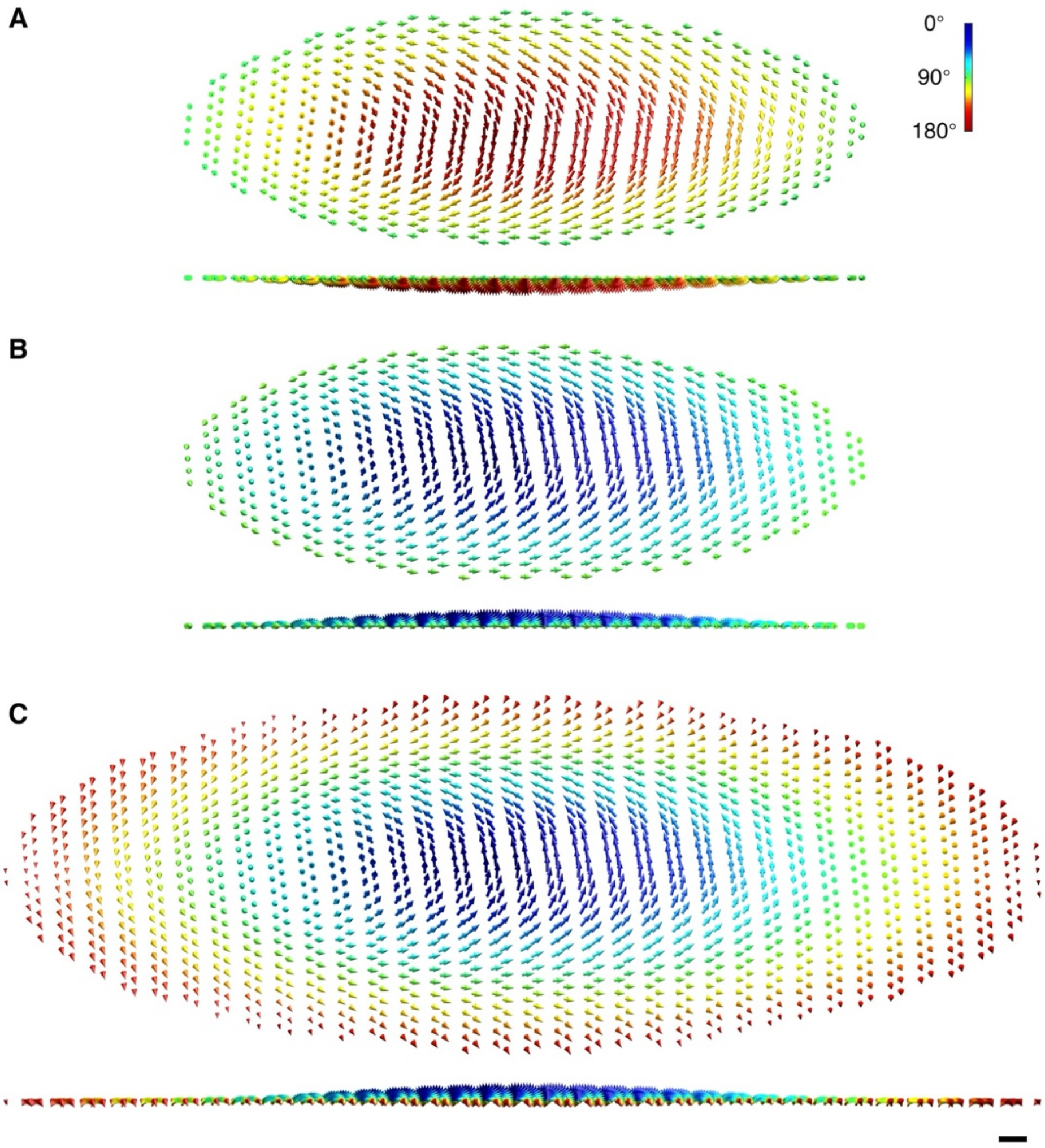

**Fig. S21. Meron-like, anti-meron-like, and anti-skyrmion-like structural textures in DFT-relaxed TBG. (A–C)** 3D vector maps of a meron-like texture (A), an anti-meron-like texture (B), and an anti-skyrmion-like texture (C), with corresponding skyrmion numbers $N_s = -0.5$, 0.5, and 1.0. The anti-skyrmion-like structural texture incorporates an anti-meron-like core at its center. Vector color indicates the polar angle relative to the xy plane. Vector magnitudes are scaled by a factor of 25 for visualization. Scale bar, 2 Å.

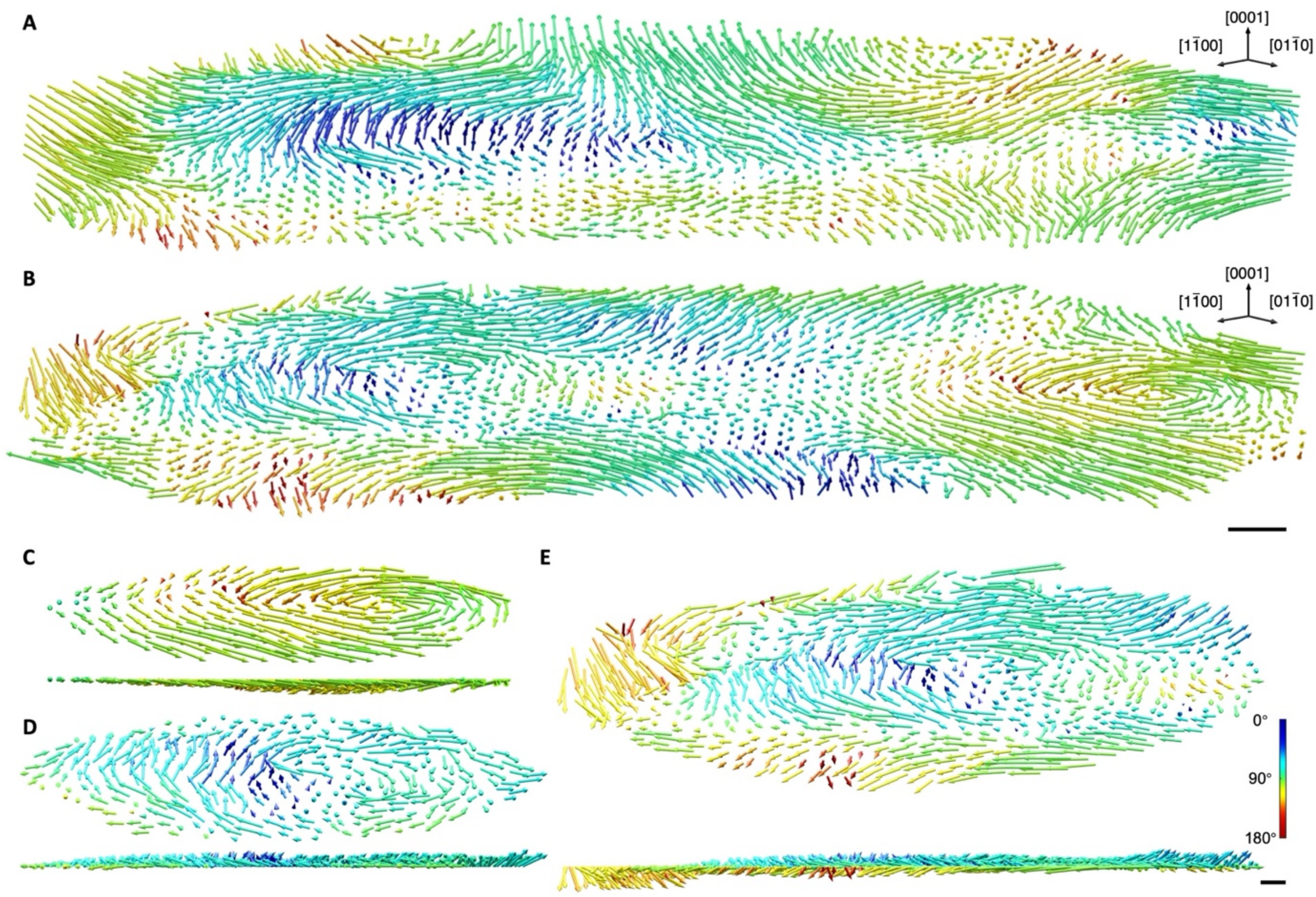

**Fig. S22**. **3D displacement vectors and chiral lattice distortions in experimental TBG. (A, B)** 3D displacement vectors in the top and bottom layers, with vector colors indicating the polar angle relative to the xy plane. (**C**, **D**) Localized vector fields corresponding to a meron-like structural texture (C) and an anti-meron-like structural texture (D). (**E**) 3D vector field of an anti-skyrmion-like structural texture. To improve displacement field precision at the expense of resolution, atomic positions were convolved with a 3D Gaussian kernel with standard deviations of 1.26 Å in x and y (half the in-plane lattice constant) and 1.71 Å in z (half the average interlayer spacing). Vector magnitudes are scaled by a factor of 25 for visualization. Scale bars, 5 Å in (B) and 2 Å in (E).

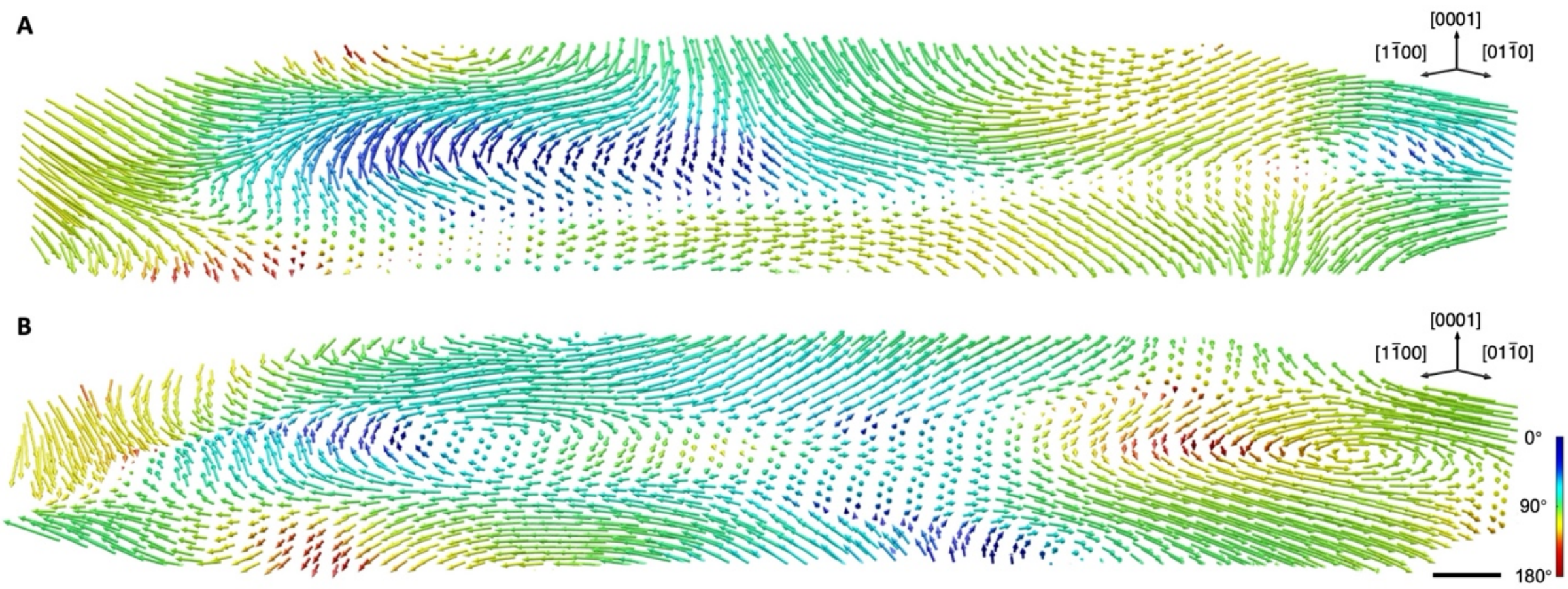

**Fig. S23.** 3D displacement vector fields from a representative leave-one-tilt-out reconstruction. (**A**, **B**) Displacement vector fields and chiral lattice distortions in the top (A) and bottom (B) graphene layers, showing good agreement with Fig. 3A and B in the main text. Vector color indicates the polar angle relative to the xy plane. Atomic positions were convolved with a 3D

Gaussian kernel, with widths matching the in-plane lattice constant and average interlayer spacing along z to enhance statistical precision. Vector magnitudes are scaled by a factor of 25 for visualization. Scale bar, 5 Å.

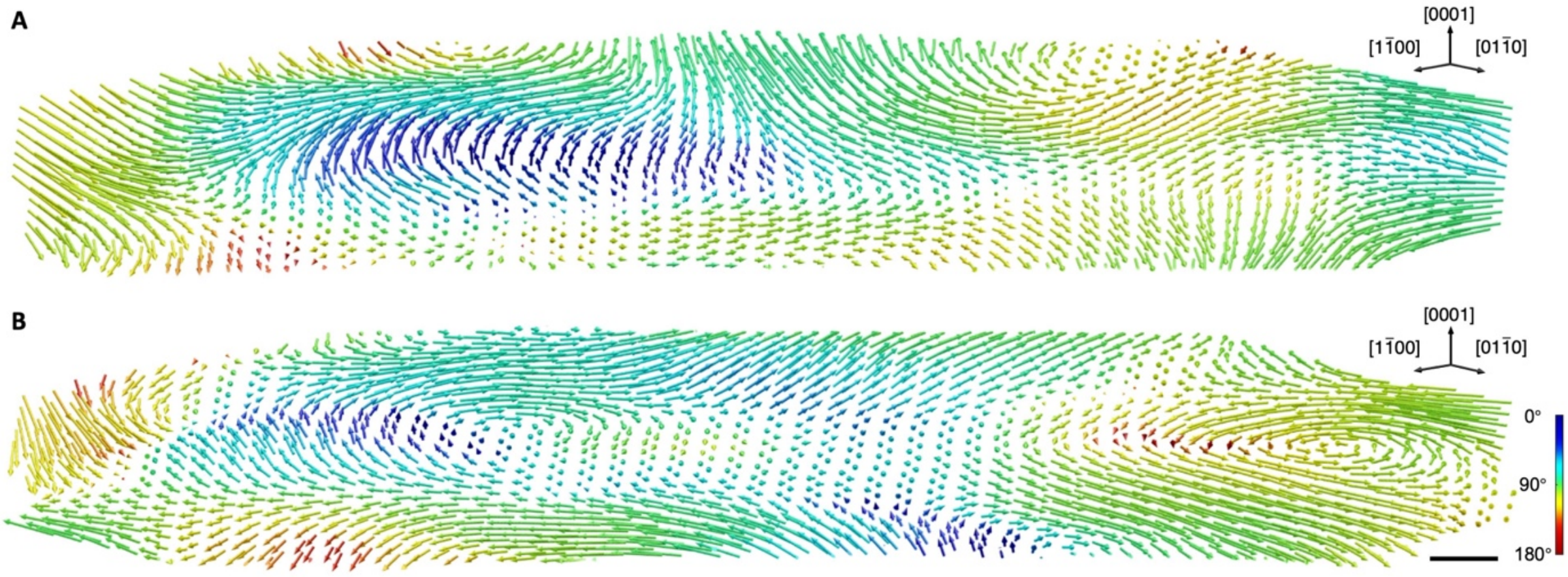

**Fig. S24. 3D displacement vector fields obtained using a pair of transient square functions as a 1D z-support.** (**A**, **B**) Displacement vector fields and chiral lattice distortions in the top (A) and bottom (B) graphene layers reconstructed with the transient square 1D z-support, showing good agreement with Fig. 3A and B obtained using a generalized Gaussian support. Vector color indicates the polar angle relative to the xy plane. Atomic positions were convolved with a 3D Gaussian kernel, with widths corresponding to the in-plane lattice constant and average interlayer spacing along z, to enhance statistical precision. Vector magnitudes are scaled by a factor of 25 for visualization. Scale bar, 5 Å.

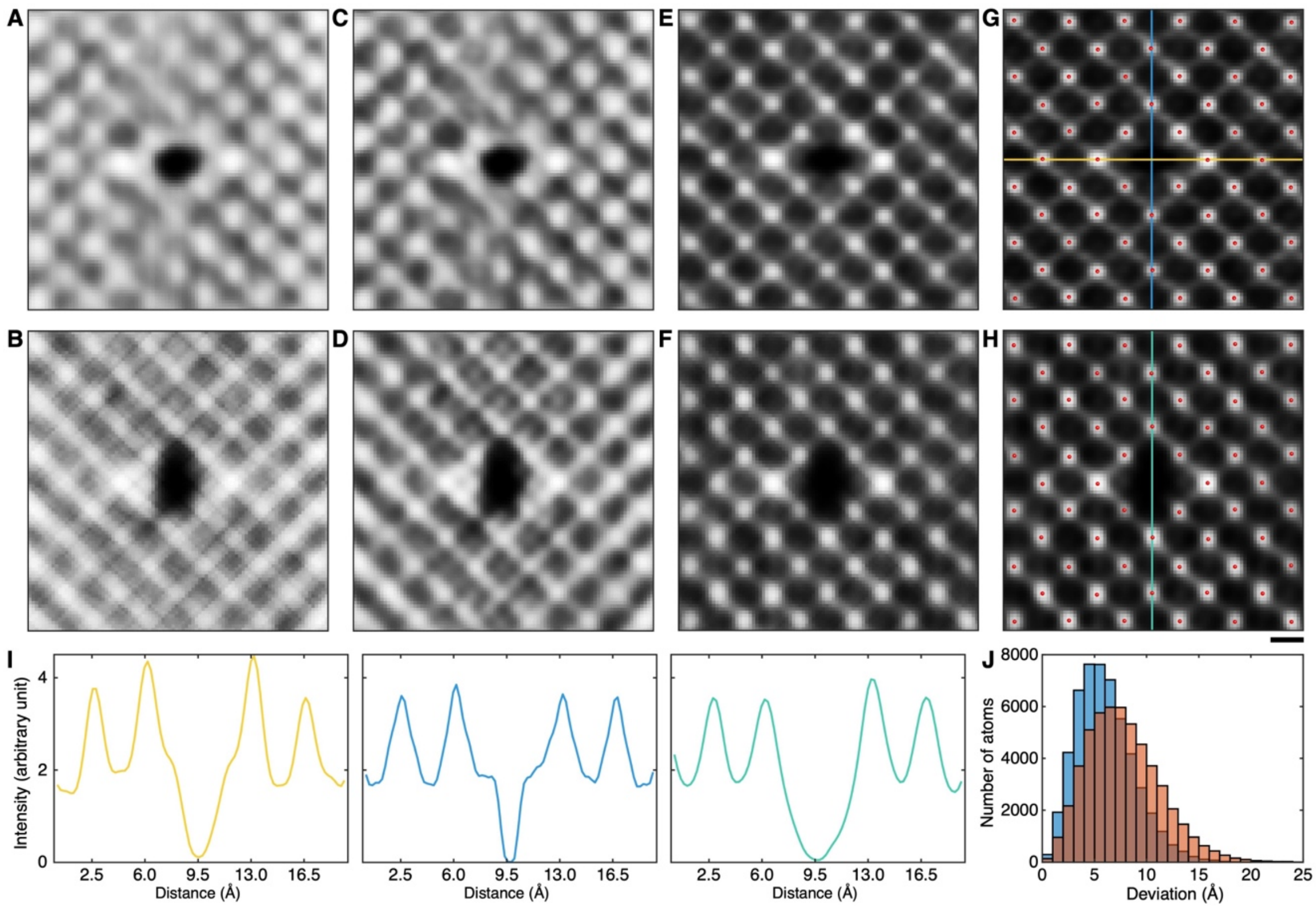

**Fig. S25. pAET simulations of an NV color center in diamond.** (**A**-**H**) 3D atomic structure of the NV center in the xy (A, C, E, G) and xz (B, D, F, H) planes. Reconstructions were performed using pAET with single-slice ptychography without (A, B) and with probe integration (C, D), as well as multislice ptychography without (E, F) and with probe integration (G, H). Scale bar, 2 Å. (**I**) Line scans along the x, y, and z directions around the NV center, obtained using multislice pAET with probe integration, demonstrating high resolution. (**J**) Histogram of deviations in 3D atomic positions. The mean RMSD values for multislice pAET with (blue) and without (orange) probe incorporation are 7 pm and 9 pm, respectively. Atom tracing in single-slice pAET was unsuccessful, emphasizing the importance of probe integration for precise atomic reconstructions.

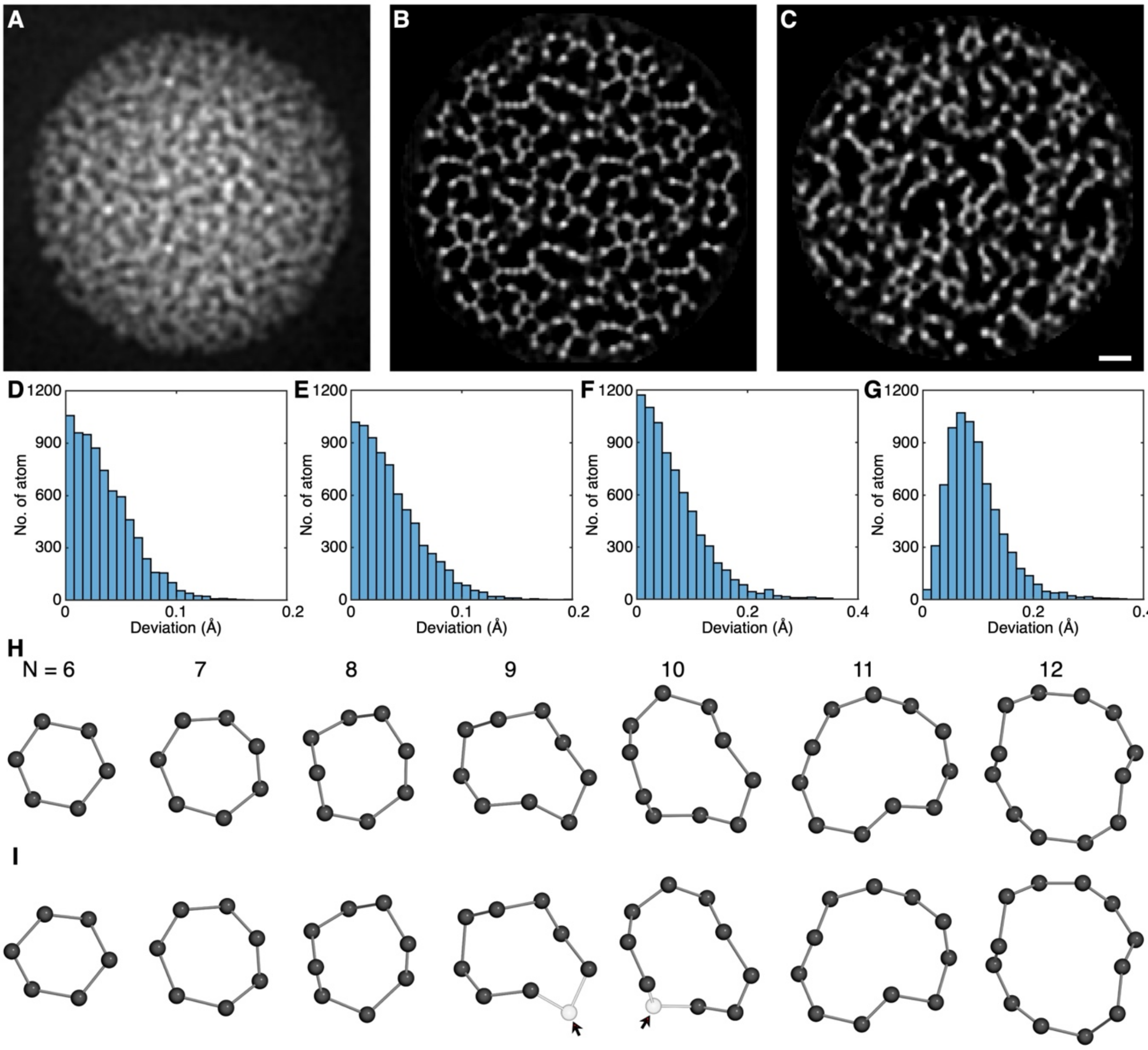

**Fig. S26. 3D atomic-resolution pAET simulation of amorphous carbon.** (**A**) Ptychographic projection at 0°, reconstructed from electron diffraction patterns with a dose of $1.5\times10^{4}$ $e^{-}$ Å$^{-2}$. (**B**, **C**) Internal slices (1.3 Å thick) from the 3D reconstruction in the xy (B) and xz (C) planes, where the z axis corresponds to the missing wedge direction. Scale bar, 5 Å. (**D**-**G**) Histograms of atomic position deviations between the reconstructed and ground-truth models along the x, y, and z axes, and overall, yielding RMSDs of 4, 5, 9, and 11 pm, respectively. A total of 99% of atoms were correctly identified. (**H**) Medium-range order motifs consisting of N-membered rings (N = 6-12) in the ground-truth atomic model. (**I**) Corresponding ring structures in the reconstructed model, with arrows indicating two unidentified atoms.

**Table S1 | Parameters for pAET data collection, reconstruction, and statistics.**

| | TBG (Exp) | TBG (Sim) | TBG (DFT+shift) | NV center in diamond (Sim) | Amorphous carbon nanoparticle |
|---|---|---|---|---|---|
| **Data Collection** | | | | | |
| Voltage (kV) | 70 | 70 | 70 | 70 | 70 |
| Detector Collection angle (mrad) | 65 | 65 | 65 | 70 | 65 |
| Convergence semi-angle (mrad) | 32 | 32 | 32 | 32 | 32 |
| Frozen phonon configurations | N/A | 64 | 32 | 16 | 8 |
| Sigma of Gaussian convolution (Å) | N/A | 0.41 | 0.41 | 0.41 | 0.41 |
| Scan step size (Å) | 0.652 | 0.652 | 0.652 | 0.4 | 0.652 |
| Number of scan positions | 230×230 | 152×152 | 192×146 | 130×400 | 100×100 |
| Number of projections | 13 | 13 | 13 | 57 | 37 |
| Pixel size (Å) | 0.095 | 0.095 | 0.095 | 0.213 | 0.340 |
| Tilt range | See fig. S7B | Same as experiment | Same as experiment | ± 70° in 2.5° increments | ± 72° in 4° increments |
| Total electron dose per projection ($e^{-}$ $Å^{-2}$) | $1.3\times10^{6}$ | $1.3\times10^{6}$ | $1.3\times10^{6}$ | $1.0\times10^{4}$ | $1.5\times10^{4}$ |
| Defocus (Å) | 0 | 0 | 0 | 0 | +13 nm |
| **Ptychography** | | | | | |
| Algorithm | Single-slice ePIE | Single-slice ePIE | Single-slice ePIE | Multi-slice Maximum likelihood | Multi-slice Maximum likelihood |
| Slice thickness(nm) | N/A | N/A | N/A | 1 | 1 |
| Number of probes | 4 | 4 | 4 | 1 | 1 |
| Number of iterations | 100 | 100 | 100 | 200 | 200 |
| **Tomography Reconstruction** | | | | | |
| Algorithm | Improved RESIRE | Improved RESIRE | Improved RESIRE | Improved RESIRE | Improved RESIRE |
| Number of iterations | 200 | 200 | 200 | 1,000 | 1,000 |
| R (%) | 11 | 8.3 | 6.1 | 2.0 | 1.6 |
| **Refinement** | | | | | |
| B' factors ($Å^2$) | | | | | |
| Type 1 atoms | 5.05 | 2.95 | 3.00 | 5.15 | 2.45 |
| Type 2 atoms | 6.83 | 3.80 | N/A | N/A | N/A |

| Statistics | | | | | |
|---|---|---|---|---|---|
| # of atoms | | | | | |
| Total | 6,652 | 6,652 | 7,222 | 52,530 | 7,514 |
| C | 6,649 | 6,649 | 7,222 | 52,529 | 7,514 |
| Si | 3 | 3 | N/A | N/A | N/A |
| N | N/A | N/A | N/A | 1* | N/A |

*Distinguishing nitrogen atoms from carbon atoms in diamond with pAET requires a higher electron dose due to the minimal difference in their atomic numbers.

## REFERENCES

1. V. Georgakilas, J. A. Perman, J. Tucek, R. Zboril, Broad Family of Carbon Nanoallotropes: Classification, Chemistry, and Applications of Fullerenes, Carbon Dots, Nanotubes, Graphene, Nanodiamonds, and Combined Superstructures. *Chem. Rev.* **115**, 4744−4822 (2015) doi:10.1021/cr500304f
2. A. K. Geim, K. S. Novoselov, The rise of graphene. *Nat. Mater.* **6**, 183−191 (2007). doi:10.1038/nmat1849
3. J. Robertson, Diamond-like amorphous carbon. *Mater. Sci. Eng. R Rep.* **37**, 129−281 (2002). doi:10.1016/S0927-796X(02)00005-0
4. J. N. Coleman, U. Khan, W. J. Blau, Y. K. Gun'ko, Small but strong: A review of the mechanical properties of carbon nanotube-polymer composites. *Carbon* **44**, 1624−1652 (2006). doi:10.1016/j.carbon.2006.02.038
5. A. A. Balandin, Thermal properties of graphene and nanostructured carbon materials. *Nat. Mater.* **10**, 569−581 (2011). doi:10.1038/nmat3064
6. I. Aharonovich, D. Englund, M. Toth, Solid-state single-photon emitters. *Nat. Photonics* **10**, 631−641 (2016). doi:10.1038/nphoton.2016.186
7. J.-C. Charlier, Defects in carbon nanotubes. *Acc. Chem. Res.* **35**, 1063−1069 (2002). doi:10.1021/ar010166k
8. M. Oh, K. P. Nuckolls, D. Wong, R. L. Lee, X. Liu, K. Watanabe, T. Taniguchi, A. Yazdani, Evidence for unconventional superconductivity in twisted bilayer graphene. *Nature* **600**, 240−245 (2021). doi:10.1038/s41586-021-04121-x
9. Y. Jiang, X. Lai, K. Watanabe, T. Taniguchi, K. Haule, J. Mao, E. Y. Andrei, Charge order and broken rotational symmetry in magic-angle twisted bilayer graphene. *Nature* **573**, 91−95 (2019). doi:10.1038/s41586-019-1460-4
10. L. J. McGilly, A. Kerelsky, N. R. Finney, K. Shapovalov, E. M. Shih, A. Ghiotto, Y. Zeng, S. L. Moore, W. Wu, Y. Bai, K. Watanabe, T. Taniguchi, M. Stengel, L. Zhou, J. Hone, X. Zhu, D. N. Basov, C. Dean, C. E. Dreyer, A. N. Pasupathy, Visualization of moiré superlattices. *Nat. Nanotechnol.* **15**, 580−584 (2020). doi:10.1038/s41565-020-0708-3
11. A. Hashimoto, K. Suenaga, A. Gloter, K. Urita, S. Iijima, Direct evidence for atomic defects in graphene layers. *Nature* **430**, 870−873 (2004). doi:10.1038/nature02817
12. C. Gómez-Navarro, J. C. Meyer, R. S. Sundaram, A. Chuvilin, S. Kurasch, M. Burghard, K. Kern, U. Kaiser, Atomic structure of reduced graphene oxide. *Nano Lett.* **10**, 1144−1148 (2010). doi:10.1021/nl9031617

13. J. C. Meyer, S. Kurasch, H. J. Park, V. Skakalova, D. Künzel, A. Gross, A. Chuvilin, G. Algara-Siller, S. Roth, T. Iwasaki, U. Starke, J. H. Smet, U. Kaiser, Experimental analysis of charge redistribution due to chemical bonding by high-resolution transmission electron microscopy. *Nat. Mater.* **10**, 209−215 (2011). doi:10.1038/nmat2941
14. J. H. Warner, E. R. Margine, M. Mukai, A. W. Robertson, F. Giustino, A. I. Kirkland, Dislocation-driven deformations in graphene. *Science* **337**, 209−212 (2012). doi:10.1126/science.1217529
15. H. Yoo, R. Engelke, S. Carr, S. Fang, K. Zhang, P. Cazeaux, S. H. Sung, R. Hovden, A. W. Tsen, T. Taniguchi, K. Watanabe, G. C. Yi, M. Kim, M. Luskin, E. B. Tadmor, E. Kaxiras, P. Kim, Atomic and electronic reconstruction at the van der Waals interface in twisted bilayer graphene. *Nat. Mater.* **18**, 448−453 (2019). doi: 10.1038/s41563-019-0346-z
16. K. Suenaga, M. Koshino, Atom-by-atom spectroscopy at graphene edge. *Nature* **468**, 1088−1090 (2010). doi:10.1038/nature09664
17. Y. Li, W. Zhou, H. Wang, L. Xie, Y. Liang, F. Wei, J. C. Idrobo, S. J. Pennycook, H. Dai, An oxygen reduction electrocatalyst based on carbon nanotube-graphene complexes. *Nat. Nanotechnol.* **7**, 394−400 (2012). doi:10.1038/nnano.2012.72
18. O. L. Krivanek, M. F. Chisholm, V. Nicolosi, T. J. Pennycook, G. J. Corbin, N. Dellby, M. F. Murfitt, C. S. Own, Z. S. Szilagyi, M. P. Oxley, S. T. Pantelides, S. J. Pennycook, Atom-by-atom structural and chemical analysis by annular dark-field electron microscopy. *Nature* **464**, 571−574 (2010). doi:10.1038/nature08879
19. J. Miao, Computational microscopy with coherent diffractive imaging and ptychography. *Nature* **637**, 281−295 (2025). doi:10.1038/s41586-024-08278-z
20. K. S. Raines, S. Salha, R. L. Sandberg, H. Jiang, J. A. Rodríguez, B. P. Fahimian, H. C. Kapteyn, J. Du, J. Miao, Three-dimensional structure determination from a single view. *Nature* **463**, 214−217 (2010). doi:10.1038/nature08705
21. A. M. Maiden, M. J. Humphry, J. M. Rodenburg, Ptychographic transmission microscopy in three dimensions using a multi-slice approach. *J. Opt. Soc. Am. A* **29**, 1606-1614 (2012). doi:10.1364/JOSAA.29.001606
22. S. Gao, P. Wang, F. Zhang, G. T. Martinez, P. D. Nellist, X. Pan, A. I. Kirkland, Electron ptychographic microscopy for three-dimensional imaging. *Nat. Commun.* **8**, 163 (2017). doi:10.1038/s41467-017-00150-1
23. Z. Chen, Y. Jiang, Y. T. Shao, M. E. Holtz, M. Odstrčil, M. Guizar-Sicairos, I. Hanke, S. Ganschow, D. G. Schlom, D. A. Muller, Electron ptychography achieves atomic-resolution limits set by lattice vibrations. *Science* **372**, 826−831 (2021). doi:10.1126/science.abg2533
24. C. M. O'Leary, H. Sha, J. Zhang, C. Su, S. Kahn, H. Jiang, A. Zettl, J. Ciston, J. Miao, Three-dimensional structure of buried heterointerfaces revealed by multislice ptychography. *Phys. Rev. Appl.* **22**, 014016 (2024). doi:10.1103/PhysRevApplied.22.014016
25. Z. Dong, M. Huo, J. Li, J. Li, P. Li, H. Sun, L. Gu, Y. Lu, M. Wang, Y. Wang, Z. Chen, Visualization of oxygen vacancies and self-doped ligand holes in $La_3Ni_2O_{7-\delta}$. *Nature* **630**, 847−852 (2024). doi:10.1038/s41586-024-07482-1
26. Z. Dong, Y. Zhang, C. Chiu, S. Lu, J. Zhang, Y. Liu, S. Liu, J. Yang, P. Yu, Y. Wang, Z. Chen, Sub-nanometer depth resolution and single dopant visualization achieved by tilt-coupled multislice electron ptychography. *Nat. Commun*. **16**, 1219 (2025). doi:10.1038/s41467-025-56499-1
27. J. Miao, P. Ercius, S. J. Billinge, 3D structures without crystals. *Science* **353**, aaf2157 (2016). doi:10.1126/science.aaf2157
28. Y. Yang, C. C. Chen, M. C. Scott, C. Ophus, R. Xu, A. Pryor, L. Wu, F. Sun, W. Theis, J. Zhou, M. Eisenbach, P. R. Kent, R. F. Sabirianov, H. Zeng, P. Ercius, J. Miao, Deciphering chemical order/disorder and material properties at the single-atom level. *Nature* **542**, 75−79 (2017). doi:10.1038/nature21042
29. X. Tian, D. S. Kim, S. Yang, C. J. Ciccarino, Y. Gong, Y. Yang, Y. Yang, B. Duschatko, Y. Yuan, P. M. Ajayan, J. C. Idrobo, P. Narang, J. Miao, Correlating the three-dimensional atomic defects and electronic properties of two-dimensional transition metal dichalcogenides. *Nat. Mater.* **19**, 867−873 (2020). doi: 10.1038/s41563-020-0636-5
30. Y. Yang, J. Zhou, F. Zhu, Y. Yuan, D. J. Chang, D. S. Kim, M. Pham, A Rana, X. Tian, Y. Yao, S. J. Osher, A. K. Schmid, L. Hu, P. Ercius, J. Miao, Determining the three-dimensional atomic structure of an amorphous solid. *Nature* **592**, 60−64 (2021). doi:10.1038/s41586-021-03354-0
31. D. J. Chang, D. S. Kim, A. Rana, X. Tian, J. Zhou, P. Ercius, J. Miao, Ptychographic atomic electron tomography: Towards three-dimensional imaging of individual light atoms in materials. *Phys. Rev*. B **102**, 174101 (2020). doi:10.1103/PhysRevB.102.174101
32. J. Lee, M. Lee, Y. Park, C. Ophus, Y. Yang, Multislice Electron Tomography Using Four-Dimensional Scanning Transmission Electron Microscopy. *Phys. Rev. Appl.* **19**, 054062 (2023). doi:10.1103/PhysRevApplied.19.054062
33. P. M. Pelz, S. M. Griffin, S. Stonemeyer, D. Popple, H. DeVyldere, P. Ercius, A. Zettl, M. C. Scott, C. Ophus, Solving complex nanostructures with ptychographic atomic electron tomography. *Nat. Commun*. **14**, 7906 (2023). doi:10.1038/s41467-023-43634-z
34. A. Romanov, M. G. Cho, M. C. Scott, P. M. Pelz, Multi-slice electron ptychographic tomography for three-dimensional phase-contrast microscopy beyond the depth of focus limits. *J. Phys. Mater.* **8**, 015005 (2025). doi:10.1088/2515-7639/ad9ad2
35. E. Y. Andrei, A. H. MacDonald, Graphene bilayers with a twist. *Nat. Mater.* **19**, 1265−1275 (2020). doi:10.1038/s41563-020-00840-0
36. K. S. Novoselov, A. Mishchenko, A. Carvalho, A. H. Castro Neto, 2D materials and van der Waals heterostructures. *Science* **353**, aac9439 (2016). doi: 10.1126/science.aac9439
37. Materials, methods and supplementary text are available as supplementary materials.
38. Y. Jiang, Z. Chen, Y. Han, P. Deb, H. Gao, S. Xie, P. Purohit, M. W. Tate, J. Park, S. M. Gruner, V. Elser, D. A. Muller, Electron ptychography of 2D materials to deep sub-ångström resolution. *Nature* **559**, 343−349 (2018). doi:10.1038/s41586-018-0298-5
39. A. M. Maiden, J. M. Rodenburg, An improved ptychographical phase retrieval algorithm for diffractive imaging. *Ultramicroscopy* **109**, 1256−1262 (2009). doi:10.1016/j.ultramic.2009.05.012

40. Z. Chen, M. Odstrčil, Y. Jiang, Y. Han, M. H. Chiu, L. J. Li, D. A. Muller, Mixed-state electron ptychography enables sub-angstrom resolution imaging with picometer precision at low dose. *Nat. Commun*. **11**, 2994 (2020). doi:10.1038/s41467-020-16688-6
41. M. Guizar-Sicairos, A. Diaz, M. Holler, M. S. Lucas, A. Menzel, R. A. Wepf, O. Bunk, Phase tomography from x-ray coherent diffractive imaging projections. *Opt. Express* **19**, 21345–21357 (2011). doi:10.1364/OE.19.021345.
42. M. Pham, Y. Yuan, A. Rana, S. Osher, J. Miao, Accurate real space iterative reconstruction (RESIRE) algorithm for tomography. *Sci. Rep.* **13**, 5624 (2023). doi:10.1038/s41598-023-31124-7
43. E. J. Kirkland, *Advanced Computing in Electron Microscopy*, 2nd ed. (Springer, New York, 2010).
44. T. Ohta, A. Bostwick, T. Seyller, K. Horn, E. Rotenberg, Controlling the electronic structure of bilayer graphene. *Science* **313**, 951−954 (2006). doi:10.1126/science.1130681
45. N. P. Kazmierczak, M. Van Winkle, C. Ophus, K. C. Bustillo, S. Carr, H. G. Brown, J. Ciston, T. Taniguchi, K. Watanabe, D. K. Bediako, Strain fields in twisted bilayer graphene. *Nat. Mater.* **20**, 956−963 (2021). doi:10.1038/s41563-021-00973-w
46. S. H. Sung, N. Schnitzer, S. Novakov, I. El Baggari, X. Luo, J. Gim, N. M. Vu, Z. Li, T. H. Brintlinger, Y. Liu, W. Lu, Y. Sun, P. B. Deotare, K. Sun, L. Zhao, L. F. Kourkoutis, J. T. Heron, R. Hovden, Two-dimensional charge order stabilized in clean polytype heterostructures. *Nat. Commun*. **13**, 7826 (2022). doi: 10.1038/s41467-021-27947-5
47. X. Z. Yu, W. Koshibae, Y. Tokunaga, K. Shibata, Y. Taguchi, N. Nagaosa, Y. Tokura, Transformation between meron and skyrmion topological spin textures in a chiral magnet. *Nature* **564**, 95−98 (2018). doi:10.1038/s41586-018-0745-3
48. S. Carr, S. Fang, E. Kaxiras, Electronic-structure methods for twisted moiré layers. *Nat. Rev. Mater.* **5**, 748−763 (2020). doi:10.1038/s41578-020-0214-0
49. A. H. Castro Neto, F. Guinea, N. M. R. Peres, K. S. Novoselov, A. K. Geim, The electronic properties of graphene. *Rev. Mod. Phys.* **81**, 109−162 (2009). doi: 10.1103/RevModPhys.81.109
50. G. Grüner, The dynamics of charge-density waves. *Rev. Mod. Phys.* **60**, 1129−1181 (1988). doi:10.1103/RevModPhys.60.1129
51. J. F. Scott, Applications of modern ferroelectrics. *Science* **315**, 954−959 (2007). doi:10.1126/science.1129564
52. Y. Yang, J. Zhou, Z. Zhao, G. Sun, S. Moniri, C. Ophus, Y. Yang, Z. Wei, Y. Yuan, C. Zhu, Y. Liu, Q. Sun, Q. Jia, H. Heinz, J. Ciston, P. Ercius, P. Sautet, Y. Huang, J. Miao, Atomic-scale identification of active sites of oxygen reduction nanocatalysts. *Nat. Catal.* **7**, 796−806 (2024). doi:10.1038/s41929-024-01175-8
53. A. Reina, X. Jia, J. Ho, D. Nezich, H. Son, V. Bulovic, M. S. Dresselhaus, J. Kong, Large area, few-layer graphene films on arbitrary substrates by chemical vapor deposition. *Nano Lett*. **9**, 30−35 (2009). doi:10.1021/nl801827v
54. A. Zobelli, A. Gloter, C. P. Ewels, G. Seifert, C. Colliex, Electron knock-on cross section of carbon and boron nitride nanotubes. *Phys. Rev. B* **75**, 245402 (2007). doi:10.1103/PhysRevB.75.245402

55. J. C. Meyer, F. Eder, S. Kurasch, V. Skákalová, J. Kotakoski, H. J. Park, S. Roth, A. Chuvilin, S. Eyhusen, G. Benner, A. V. Krasheninnikov, U. Kaiser, Accurate measurement of electron-beam-induced displacement cross sections for single-layer graphene. *Phys. Rev. Lett.* **108**, 196102 (2012). doi:10.1103/PhysRevLett.108.196102
56. S. Cao, P. Kok, P. Li, A. M. Maiden, J. M. Rodenburg, Modal decomposition of a propagating matter wave via electron ptychography. *Phys. Rev. A* **94**, 063621 (2016). doi:10.1103/PhysRevA.94.063621
57. M. C. Scott, C. C. Chen, M. Mecklenburg, C. Zhu, R. Xu, P. Ercius, U. Dahmen, B. C. Regan, J. Miao, Electron tomography at 2.4-ångström resolution. *Nature* **483**, 444−447 (2012). doi:10.1038/nature10934
58. S. Nadarajah, A generalized normal distribution. *J. Appl. Stat.* **32**, 685−694 (2005). doi:10.1080/02664760500079464
59. S. S. Rogers, T. A. Waigh, X. Zhao, J. R. Lu, Precise particle tracking against a complicated background: polynomial fitting with Gaussian weight. *Phys. Biol.* **4**, 220−227 (2007). doi:10.1088/1478-3975/4/3/008
60. J. Madsen, T. Susi, The abTEM code: Transmission electron microscopy from first principles. *Open Res. Eur.* **1**, 24 (2021). doi:10.12688/openreseurope.13015.2
61. Y. Tserkovnyak, J. Zou, S. K. Kim, S. Takei, Quantum hydrodynamics of spin winding. *Phys. Rev. B* **102**, 224433 (2020). doi:10.1103/PhysRevB.102.224433
62. A. Rana, C. T. Liao, E. Iacocca, J. Zou, M. Pham, X. Lu, E. C. Subramanian, Y. H. Lo, S. A. Ryan, C. S. Bevis, R. M. Karl Jr, A. J. Glaid, J. Rable, P. Mahale, J. Hirst, T. Ostler, W. Liu, C. M. O'Leary, Y. S. Yu, K. Bustillo, H. Ohldag, D. A. Shapiro, S. Yazdi, T. E. Mallouk, S. J. Osher, H. C. Kapteyn, V. H. Crespi, J. V. Badding, Y. Tserkovnyak, M. M. Murnane, J. Miao, Three-dimensional topological magnetic monopoles and their interactions in a ferromagnetic meta-lattice. *Nat. Nanotechnol.* **18**, 227−232 (2023). doi:10.1038/s41565-022-01311-0
63. K. Uchida, S. Furuya, J.-I. Iwata, A. Oshiyama, Atomic corrugation and electron localization due to Moiré patterns in twisted bilayer graphenes. *Phys. Rev. B* **90**, 155451 (2014). doi:10.1103/PhysRevB.90.155451
64. P. Lucignano, D. Alfè, V. Cataudella, D. Ninno, G. Cantele, Crucial role of atomic corrugation on the flat bands and energy gaps of twisted bilayer graphene at the magic angle $\theta \sim 1.08°$. *Phys. Rev. B* **99**, 195419 (2019). doi:10.1103/PhysRevB.99.195419.
65. G. Kresse, J. Furthmüller, Efficient iterative schemes for ab initio total-energy calculations using a plane-wave basis set. *Phys Rev B* **54**, 11169−11186 (1996). doi:10.1103/PhysRevB.54.11169
66. I. Hamada, van der Waals density functional made accurate. *Phys Rev B* **89**, 121103 (2014). doi:10.1103/PhysRevB.89.121103
67. C. P. Herrero, R. Ramírez, Elastic properties and mechanical stability of bilayer graphene: molecular dynamics simulations. *Eur. Phys. J. B* **96**, 147 (2023). doi:10.1140/epjb/s10051-023-00616-w
68. M. Shishkin, G. Kresse, Self-consistent *GW* calculations for semiconductors and insulators. *Phys. Rev. B* **75**, 235102 (2007). doi:10.1103/PhysRevB.75.235102.
69. P. E. Blöchl, Projector augmented-wave method. *Phys Rev B* **50**, 17953−17979 (1994). doi:10.1103/PhysRevB.50.17953

70. G. Kresse, D. Joubert, From ultrasoft pseudopotentials to the projector augmented-wave method. *Phys. Rev. B* **59**, 1758 (1999). doi:10.1103/PhysRevB.59.1758
71. M. Odstrčil, A. Menzel, M. Guizar-Sicairos, Iterative least-squares solver for generalized maximum-likelihood ptychography. *Opt. Express* **26**, 3108−3123 (2018). doi:10.1364/OE.26.003108
72. K. Wakonig, H. C. Stadler, M Odstrčil, E. H. R. Tsai, A. Diaz, M. Holler, I. Usov, J. Raabe, A. Menzel, M. Guizar-Sicairos, PtychoShelves, a versatile high-level framework for high-performance analysis of ptychographic data. *J. Appl. Crystallogr.* **53**, 574−586 (2020). doi:10.1107/S1600576720001776
73. E. H. R. Tsai, I. Usov, A. Diaz, A. Menzel, M. Guizar-Sicairos, X-ray ptychography with extended depth of field. *Opt. Express* **24**, 29089−29108 (2016). doi:10.1364/OE.24.029089
74. J. Tersoff, Empirical interatomic potential for carbon, with applications to amorphous carbon, *Phys. Rev. Lett*. **61**, 2879 (1988). doi:10.1103/PhysRevLett.61.2879